\documentclass[vecphys]{svmult}
\authorrunning{S. Phang, et al.}
\titlerunning{Boundary Integral Equation for $\mathcal{PT}$-symmetric Resonator}

\usepackage{makeidx}         

\usepackage{graphicx}\usepackage{epstopdf} 
\usepackage{overpic} 
\usepackage{rotating}

\usepackage{amsmath,esint}\usepackage{amssymb} \usepackage{mathrsfs} \usepackage{mathtools}  
\usepackage{bm} 

\usepackage{multicol}        
\usepackage{cite}            
\usepackage[perpage]{footmisc}

\usepackage{xcolor,calc, blindtext} 

\usepackage{booktabs}  

\usepackage{comment} 
	\definecolor{shadecolor}{gray}{0.80}
	\excludecomment{comment}  

\makeindex             

\newcommand{\e}{{\rm e}}
\renewcommand{\d}{{\rm d}}
\newcommand{\x}{{\bf x}}

\newcommand{\dydxv}[2]{ {{\partial #1} \over {\partial #2}} }

\newcommand{\PT}{{$\mathcal{PT}$}}

\newcommand{\posx}{{\bf x}}

\begin{document}
\setcounter{chapter}{6}

\title{Theory and Numerical Modelling of Parity-Time Symmetric Structures in Photonics: Boundary Integral Equation for Coupled Microresonator Structures}
\author{S. Phang\textsuperscript{1,2,*} \and A. Vukovic\textsuperscript{2} \and G. Gradoni\textsuperscript{1,2} \and P. D. Sewell\textsuperscript{2}  \and T. M. Benson\textsuperscript{2} \and S. C. Creagh\textsuperscript{1}}
\institute{
	Wave Modelling Research Group - School of Mathematical Sciences, University of Nottingham, United Kingdom
	\and George Green Institute for Electromagnetics Research, University of Nottingham, United Kingdom  \\ 
	\textsuperscript{*} Corresponding author - \texttt{sendy.phang@nottingham.ac.uk}
}

\maketitle

\begin{abstract}
The spectral behaviour and the real-time operation of Parity-Time (\PT) symmetric coupled resonators are investigated. A Boundary Integral Equation (BIE) model is developed to study these structures in the frequency domain. The impact of realistic gain/loss material properties on the operation of the \PT-symmetric coupled resonators is also investigated using the time-domain Transmission-Line Modelling (TLM) method. {The BIE method is also used to study the behaviour of an array of PT-microresonator photonic molecules.}    
\end{abstract}

\noindent
 
\section{Introduction}
In Chapter 6, the distinctive features and properties and great potential of Parity-Time (\PT) symmetric structure \index{\PT-symmetric structure} in photonics were introduced. The focus, there, was on one-dimensional structures that were based on Bragg gratings. The study of these structures was extended to include non-ideal material properties via a time-domain Transmission-Line Modelling (TLM) \index{Transmission-Line Modelling (TLM)} method. In this chapter, the impact of the realistic gain/loss parameter, which was introduced in the previous chapter, on the spectral properties of a \PT-symmetric system based on two coupled microresonators is studied. A semi-analytical model based on a Boundary Integral Equation (BIE) \index{Boundary Integral Equation (BIE)} is developed for this study in the frequency domain. On the other hand, the real-time operation will be subsequently studied by using an extended time-domain numerical Transmission-Line Modelling (TLM) method in two-dimensions; the method was introduced in the previous chapter for one-dimensional problems.

The chapter starts by extending the one-dimensional TLM model incorporating the dispersive gain/loss model, which was described in the previous chapter, to two-dimensions. {The subsequent section will detail the development of the BIE model for the \PT -coupled resonator\index{\PT-coupled resonator}. This is followed by a study of the influence of material gain/loss on the spectral properties of the \PT-coupled resonator, and an investigation of the real-time operation of the \PT-coupled resonator using the TLM method. Finally, the behaviour of an array of \PT-microresonator photonic molecules is studied using the BIE method.}   

It is noted here that although most of the content of this chapter is self-contained, it is the second of two chapters in this book covering the theory and numerical modelling of Parity-Time symmetric structures in photonics. Some introductory discussion on Parity-Time (\PT) symmetry and details of the digital filter system for modelling dispersive gain/loss materials within the Transmission-Line Modelling (TLM) method which is covered in Chapter 6 of this book, will not be reproduced again in the current chapter.  

\section{The Transmission-Line Modelling Method for Dispersive Gain (or Loss) in Two-Dimension} \label{sec:2dtlm}
This section briefly reviews the time-domain Transmission-Line Modelling (TLM) method in two-dimensions. The TLM method presented here is the alternative formulation based on the bilinear transform implementation. This alternative form of TLM is suitable in the modelling of active material with dispersive and non-linear properties. This section further shows the extension of the digital filter for gain/loss material model developed in Chapter 6. 
\subsection{TLM Formalism in 2D Domain} 
Consider Maxwell's equations \index{Maxwell's equations} defined in a Cartesian coordinate system as,
\begin{align}
&\begin{cases}
(\nabla\times \bm{H}) \cdot \hat{x} &= J_{ex} + \dfrac{\partial D_x}{\partial t} \\
(\nabla\times \bm{H}) \cdot \hat{y} &= J_{ey} + \dfrac{\partial D_y}{\partial t} \\
(\nabla\times \bm{H}) \cdot \hat{z} &= J_{ez} + \dfrac{\partial D_z}{\partial t} 
\end{cases} 
\label{eq:expandmax1}
\\	
&\begin{cases}
(\nabla\times \bm{E}) \cdot \hat{x} &=  -\dfrac{\partial B_x}{\partial t} \\
(\nabla\times \bm{E}) \cdot \hat{y}  &=  -\dfrac{\partial B_y}{\partial t} \\
(\nabla\times \bm{E}) \cdot \hat{z} &=  -\dfrac{\partial B_z}{\partial t} 
\end{cases} 	
\label{eq:expandmax2}
\end{align}   
where $\hat{x}$, $\hat{y}$ and $\hat{z}$ are unit vector elements in the $x$, $y$ and $z$ direction and $(\text{ }\cdot \text{  })$ denotes the vector product.

In two-dimensions (2D), the electromagnetic fields ($E_{x,y,z}$ and $H_{x,y,z}$) are invariant in one direction. For consistency, it is taken to be the $z$-direction hence,
\begin{align}
\frac{\partial }{\partial z} \equiv 0 
\label{eq:zinvar}
\end{align} 
Implementation of the condition Eq. (\ref{eq:zinvar}) within Eq. (\ref{eq:expandmax1}) and Eq. (\ref{eq:expandmax2}) leads to two sets of uncoupled Maxwell's equations associated to $E_{z}$ or $H_{z}$. As such an $E$-type wave has $E_z$ as the primary field component and an $H$-type wave has  $H_z$ as the primary field component  throughout this chapter. 

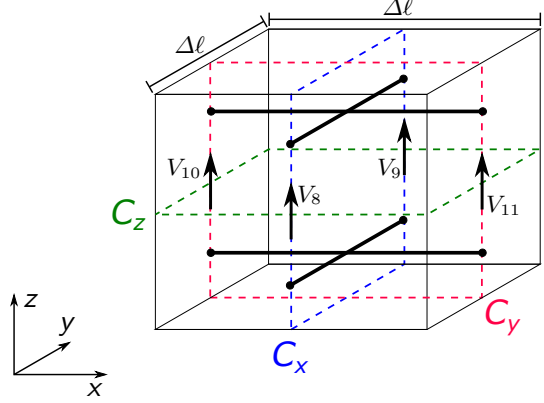
\begin{figure}[t]
	\begin{overpic}[width=0.6\textwidth,tics=5]{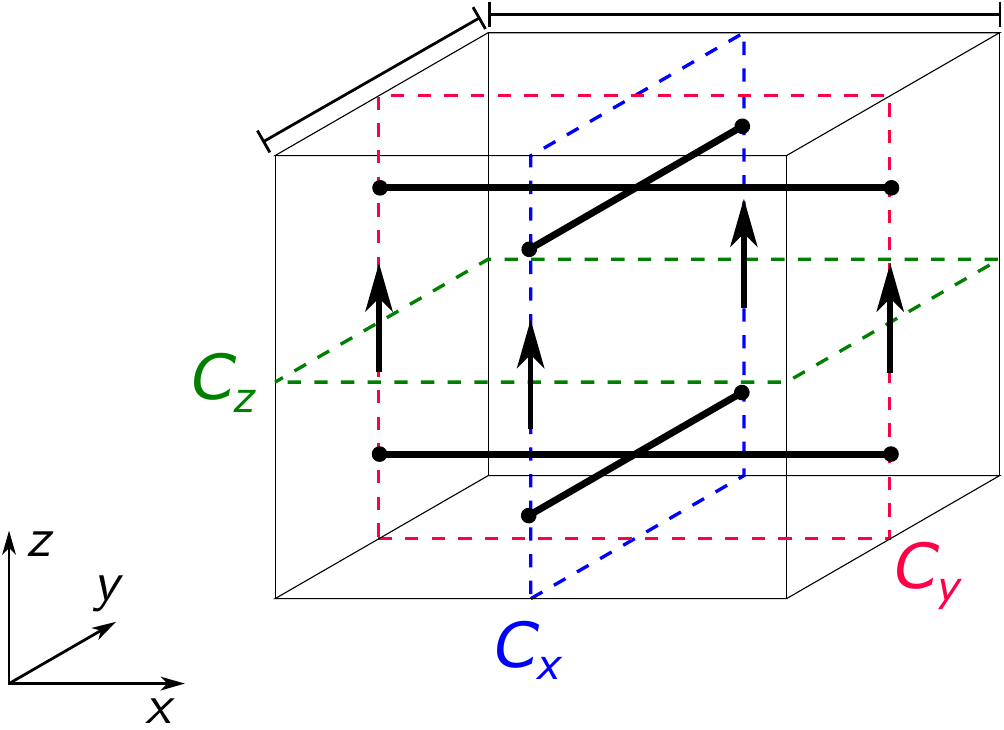}
		\put(54,37){$V_8$} \put(69,42){$V_9$} \put(29.5,42){$V_{10}$} \put(89.5,35){$V_{11}$}
		\put(31,65){$\Delta \ell$} \put(70,72){$\Delta \ell$}
		\end{overpic}
		\centering
		\caption[Schematic of 2D-TLM nodes for an $E$-type wave.]{{Schematic of 2D-TLM nodes for an $E$-type wave. A structured TLM meshing paradigm is considered, i.e. rectangle based meshing. Three different integration contours $C_x$, $C_y$ and $C_z$ are denoted by their normal axes.}}
		\label{fig:08_tlmshuntillust}
\end{figure}

Maxwell's equations for $E$-type waves, are given by, 
\begin{align}
\begin{cases}
(\nabla\times \bm{E}) \cdot \hat{x} &=  -\dfrac{\partial B_x}{\partial t} \\
(\nabla\times \bm{E}) \cdot \hat{y} &=  -\dfrac{\partial B_y}{\partial t}  \\
(\nabla\times \bm{H}) \cdot \hat{z} &= J_{ez} + \dfrac{\partial D_z}{\partial t}  
\end{cases} 
\label{eq:2dmax}
\end{align}
Upon substituting the constitutive relations for isotropic, homogeneous and non-magnetic material, Eq. (\ref{eq:2dmax}) can also be expressed as,
\begin{align}
\begin{cases}
(\nabla\times \bm{E}) \cdot \hat{x} &=  -\mu_0\dfrac{\partial H_x}{\partial t}  \\
(\nabla\times \bm{E}) \cdot \hat{y} &=  -\mu_0\dfrac{\partial H_y}{\partial t}  \\
(\nabla\times \bm{H}) \cdot \hat{z} &= \sigma_e\ast E_z + \varepsilon_0 \dfrac{\partial E_z}{\partial t} + \dfrac{\partial P_{ez}}{\partial t} 
\end{cases}
\label{eq:2dmax1}
\end{align}
In Eq. (\ref{eq:2dmax1}), non-magnetic material has been assumed, i.e. $\mu=\mu_0$ and $*$ is used to denote a convolution operation. Upon application of the transmission-line theory, which was described in the previous chapter, a 4-port \textit{shunt} transmission line \index{shunt transmission line}, which is obtained by concatenating the 1D transmission lines (see Fig. \ref{fig:08_tlmshuntillust}), can be utilised to model the $E$-type wave propagation \index{$E$-type wave propagation}. Figure \ref{fig:08_tlmshuntillust} depicts the schematic of a single 2D-TLM shunt node in a Cartesian system along with the voltage at each of its 4 ports. 

It can be seen from Fig. \ref{fig:08_tlmshuntillust} that the shunt node has 4 ports which, for consistency with \cite{Paul2001,Paul1998,Paul1999b,Paul1999a,Paul2002,Christopoulos1995}, are called ports 8, 9, 10 and 11. Hence the corresponding voltages at these ports are denoted as $V_8$, $V_9$, $V_{10}$ and $V_{11}$. Moreover the differential operators $\nabla$ and $\partial/\partial t $ can be normalised by $\Delta \ell$ and $\Delta t$ as, 
\begin{align}
\begin{cases}
\dfrac{1}{\Delta \ell} (\bar{\nabla}\times \bm{E}) \cdot \hat{x} &=  -\mu_0 \dfrac{1}{\Delta t} \dfrac{\partial H_x}{\partial T}  \\
\dfrac{1}{\Delta \ell} (\bar{\nabla}\times \bm{E}) \cdot \hat{y} &=  -\mu_0 \dfrac{1}{\Delta t} \dfrac{\partial H_y}{\partial T} \\
\dfrac{1}{\Delta \ell} (\bar{\nabla}\times \bm{H}) \cdot \hat{z} &= \sigma_e\ast E_z + \varepsilon_0  \dfrac{1}{\Delta t}  \dfrac{\partial E_z}{\partial T} +  \dfrac{1}{\Delta t}  \dfrac{\partial P_{ez}}{\partial T} 
\end{cases}
\label{eq:tlmderi1}
\end{align}

where, $T\equiv t\Delta t$ and $\bar{\nabla}\equiv \nabla\Delta \ell$ are the (dimensionless) normalised parameters, $\Delta \ell$ denotes the length of the side of an unit cell and $\Delta t$ denote the time step of the TLM calculation. The relation between $\Delta \ell$ and $\Delta t$ is discussed below. Furthermore, by implementing the field-circuit equivalences \index{field-circuit equivalences} (See Table 6.2 in Chapter 6), the Maxwell's equations \index{Maxwell's equations} (\ref{eq:tlmderi1}) can be shown in the circuit form as, 
\begin{align}
\begin{cases}
-\dfrac{1}{\Delta \ell^2} (\bar{\nabla}\times \bm{V}) \cdot \hat{x} &= 
\mu_0 \dfrac{1}{Z_\text{TL}\Delta \ell\Delta t} \dfrac{\partial i_x}{\partial T}   \\
-\dfrac{1}{\Delta \ell^2} (\bar{\nabla}\times \bm{V}) \cdot \hat{y} &=  
\mu_0 \dfrac{1}{Z_\text{TL}\Delta \ell\Delta t} \dfrac{\partial i_y}{\partial T}  \\
\dfrac{1}{Z_\text{TL}\Delta \ell^2} (\bar{\nabla}\times \bm{i}) \cdot \hat{z} &= 
\dfrac{\sigma_e}{\Delta \ell}\ast V_z +  \dfrac{\varepsilon_0 }{\Delta \ell\Delta T}  \dfrac{\partial V_z}{\partial T} +  \dfrac{\varepsilon_0}{\Delta \ell\Delta t}  \dfrac{\partial p_{ez}}{\partial T}  
\end{cases}
\label{eq:tlmderi2}
\end{align}            
Since a two-dimensional TLM model based on structured meshing is developed in this chapter, i.e. rectangular based spatial discretisation in 2D, where $\Delta x=\Delta y=\Delta \ell$, it is customary to define the transmission-line impedance $Z_\text{TL}$ and the unit transit time $\Delta t$ to correspond to the properties of wave propagation in free-space with a $45^{\circ}$ angle \cite{Paul2001,Paul1998,Paul1999b,Paul1999a,Paul2002,Christopoulos1995,Janyani2005,Balanis2012} as, 
\begin{align}
&Z_\text{TL} = \frac{\eta_0}{\sin 45^\circ} = \sqrt{2}\eta_0, \\ 
&v_\text{TL} = \dfrac{\Delta \ell}{\Delta t} = \dfrac{c_0}{\cos 45^\circ} =\sqrt{2}c_0, 
\label{eq:2tlmcond}
\end{align}  
where $v_\text{TL}$ denotes the velocity of voltage pulse propagation between TLM nodes and  $c_0=1/\sqrt{\varepsilon_0\mu_0}$ and $\eta_0=\sqrt{\mu_0/\varepsilon_0}$ respectively denote the free-space speed of light and the free-space wave impedance of a normal wave propagation. Substituting  Eq. (\ref{eq:2tlmcond}) into Eq. (\ref{eq:tlmderi2}), yields
\begin{align}
\begin{cases}
- (\bar{\nabla}\times \bm{V}) \cdot \hat{x} &= \dfrac{\partial i_x}{\partial T}   \\
- (\bar{\nabla}\times \bm{V}) \cdot \hat{y} &= \dfrac{\partial i_y}{\partial T}  \\
(\bar{\nabla}\times \bm{i}) \cdot \hat{z} &= g_e\ast V_z + 2 \dfrac{\partial V_z}{\partial T} +  2 \dfrac{\partial p_{ez}}{\partial T}  
\end{cases}
\label{eq:tlmderi3}
\end{align}
Using Stoke's theorem to solve the curl operations on the contours $C_x$, $C_y$ and $C_z$ indicated in Fig. \ref{fig:08_tlmshuntillust}, leads to 
\begin{align}
\begin{cases}
-(V_9-V_8)  &= \dfrac{\partial i_x}{\partial T}   \\
-(V_{10}-V_{11})  &= \dfrac{\partial i_y}{\partial T}  \\
(V_8 + V_9 + V_{10}+V_{11})  &= g_e\ast V_z + 2 \dfrac{\partial V_z}{\partial T} +  2 \dfrac{\partial p_{ez}}{\partial T}  
\end{cases}
\label{eq:tlmderi4}
\end{align}
After transforming the normalised time derivative to the Laplace domain \index{Laplace domain},
\begin{align}
\begin{cases}
-(V_9-V_8)  &= \bar{s} i_x  \\
-(V_{10}-V_{11})  &= \bar{s}i_y  \\
(V_8 + V_9 + V_{10}+V_{11})  &= g_e V_z + 2\bar{s}V_z +  2\bar{s}p_{ez}   
\end{cases}
\label{eq:tlmderi5}
\end{align}
Utilising the travelling-wave format \index{travelling-wave format} \cite{Paul2001,Paul1998,Paul1999b,Paul1999a,Paul2002} of the port voltage, Eq. (\ref{eq:tlmderi4}) can be expressed as, 
\begin{align}
\begin{cases}
-2(V^i_9-V^i_8)  &= 2 i_x \\
-2(V^i_{10}-V^i_{11})  &= 2i_y \\
2(V^i_8 + V^i_9 + V^i_{10}+V^i_{11})  &= g_e V_z + 4V_z +  2\bar{s}p_{ez}  
\end{cases}
\label{eq:tlmderi6}
\end{align}
Equations (\ref{eq:tlmderi6}) are the governing equations for the TLM nodal voltage calculation which are ready for material implementation. Subsequently, after calculating the nodal field values $i_x$, $i_y$ and $V_z$, the new scattered voltage impulses in the condensed TLM nodes \index{condensed TLM nodes} can be obtained as \cite{Paul2001,Paul1998,Paul1999b,Paul1999a,Paul2002}, 
\begin{align}
\begin{cases}
V_8^r = V_z - i_x -V_9^i \\
V_9^r = V_z + i_x -V_8^i \\
V_{10}^r = V_z + i_y - V_{11}^i \\
V_{11}^r = V_z - i_y - V_{10}^i  
\end{cases}
\label{eq:2dtlmscat}
\end{align}
In this present work, a simple matching boundary condition \index{boundary condition} is implemented which gives good approximation of the radiating boundary condition \cite{Paul2001,Paul1998,Paul1999b,Paul1999a,Paul2002,Christopoulos1995,Janyani2005}. This is accomplished by matching the impedance of the \textit{modelled} material with the impedance of the transmission-line \cite{Paul2001,Paul1998,Paul1999b,Paul1999a,Paul2002,Christopoulos1995,Janyani2005}. The reflected wave for a matched boundary is given by,  
\begin{align}
V^r_{\text{on boundary}} = \Gamma V^i_{\text{on boundary}} 
\end{align}
where $\Gamma$ is the reflection coefficient given as,
\begin{align}
\Gamma = \dfrac{Z_\text{material}-Z_\text{TL}}{Z_\text{material}+Z_\text{TL}} 
\end{align}
and the impedance of the material is related to the refractive index of the material, i.e. $Z_\text{material}=\eta_0/n_\text{material}$. 

\subsection{TLM Shunt Node Model for Realistic Gain Medium}
\label{sec:tlm}
In this subsection, the TLM shunt node \index{shunt transmission line} model is developed to model the realistic dispersive and saturable gain/loss medium which was previously implemented in the 1D-TLM nodes. It follows from the discussion of the homogeneously broadened gain/loss medium in the previous chapter, the material permittivity of a dispersive gain/loss material model and no-saturation is assumed $\mathbb{S}=1$ is given by, 
\begin{align}
\bar{\varepsilon}(\omega) = \varepsilon_\infty -j \frac{\sigma_0}{2\varepsilon_0\omega}\left( 
\frac{1}{1+j(\omega+\omega_\sigma)\tau} + 
\frac{1}{1+j(\omega-\omega_\sigma)\tau}
\right) 
\label{eq:gainloss}
\end{align}
where $\varepsilon_\infty$ denotes the permittivity at infinity, $\omega_\sigma$ denotes the atomic transitional frequency, $\tau$ is the dipole relaxation time and $\sigma_0$ is related to the conductivity peak value that is set by pumping level at $\omega_\sigma$. For the detail of the physical meaning of these parameter, we refer the reader to the previous chapter.    

By performing the $\mathcal{Z}$-bilinear transformation \index{$\mathcal{Z}$-bilinear transformation} on the normalised Laplace \index{Laplace domain} variable $\bar{s}$, the frequency domain Eq. (\ref{eq:tlmderi6}) can be expressed in terms of the time-delayed voltage pulses as,  
\begin{align}
i_x^i  &= -i_x \label{eq:tlmderi7a} \\
i_y^i  &= -i_y  \label{eq:tlmderi7b}\\
2V_z^i  &= (4+g_e)V_z +  2\left(2\dfrac{1-z^{-1}}{1+z^{-1}}\right)\chi_e V_z   
\label{eq:tlmderi7}
\end{align}   
where for convenience in Eq. (\ref{eq:tlmderi7}) the incoming pulses have been renamed as $i_x^i$, $i_y^i$ and $V_z^i$ and are given by 
\begin{align}
i_x^i &= V^i_9-V^i_8 \\
i_y^i &= V^i_{10}-V^i_{11} \\
V_z^i &= V^i_8 + V^i_9 + V^i_{10}+V^i_{11}
\end{align}  
{It can be observed} from Eq. (\ref{eq:tlmderi7a}) and Eq. (\ref{eq:tlmderi7b}) that the TLM model for $E$-type waves \index{$E$-type wave propagation} and a non-magnetic material has a simple nodal calculation for the transverse field components ($i_x$ and $i_y$). The material parameters responsible for dielectric modelling $\chi_e$ and gain/loss $g_e$ are only found in Eq. (\ref{eq:tlmderi7}) which is responsible for the calculation of electric field $V_z$. Thus for clarity, consider (\ref{eq:tlmderi7}) which after multiplying both sides by $(1+z^{-1})$ and some rearrangement, yields      
\begin{align}
(1+z^{-1})(2V_z^i -4V_z) &= (1+z^{-1})g_eV_z +  4(1-z^{-1})\chi_e V_z
\label{eq:2dtlmderi8}
\end{align}
By substituting the digital filter \index{digital filter} for conductivity, Eq. (6.76) from Chapter 6, into Eq. (\ref{eq:2dtlmderi8}) and after some rearrangement, Eq. (\ref{eq:2dtlmderi8}) becomes
\begin{align}
2V_z^i + z^{-1} S_{ez} = K_{e2}V_z
\label{eq:2dtlmgain}
\end{align}
where the accumulated delayed variable $S_{ez}$,
\begin{align}
\begin{aligned}
S_{ez} &= 2V_z^i + K_{e1} V_z + S_{ecz}, \\
S_{ecz} &= -\bar{g}_e V_z 
\end{aligned}
\label{eq:2dupdate}
\end{align}
and the constants $K_{e1}$ and $K_{e2}$ are given by, 
\begin{align}
K_{e1} &= -(4+g_{e1}-4\chi_e)  \\
K_{e2} &= 4+g_{e0}+4\chi_e 
\end{align}
It is important to note that the constants $g_{e0}$ and $g_{e1}$ are the same as the ones given in the previous chapter and are reproduced below, 
\begin{align}
g_{e0} &= g_s \left( \frac{K_3}{K_6}\right)  \\
g_{e1} &= 0  \\
\bar{g}_e(z) &= \frac{b_0 + z^{-1}b_1 + z^{-2}b_2}{1 - z^{-1}(-a_1) - z^{-2}(-a_2)} 
\label{eq:gaincomponents1}
\end{align}
Details of the digital filter \index{digital filter} design of the conductivity are given in the previous chapter. Moreover, the updating scheme for the conductivity $S_{ecz}$ is as illustrated in Fig. 6.8(b) in Chapter 6. In summary the nodal calculation in the presence of a gain medium is comprised of Eq. (\ref{eq:tlmderi7a}), Eq. (\ref{eq:tlmderi7b}) and Eq. (\ref{eq:2dtlmgain}) which are subsequently followed by the updating scheme Eq. (\ref{eq:2dupdate}). 

\section{Parity-Time (\PT) Symmetric Coupled Resonators}
\label{sec:ptcoup}
\begin{figure}[t]
	\begin{overpic}[width=0.55\textwidth,tics=5]{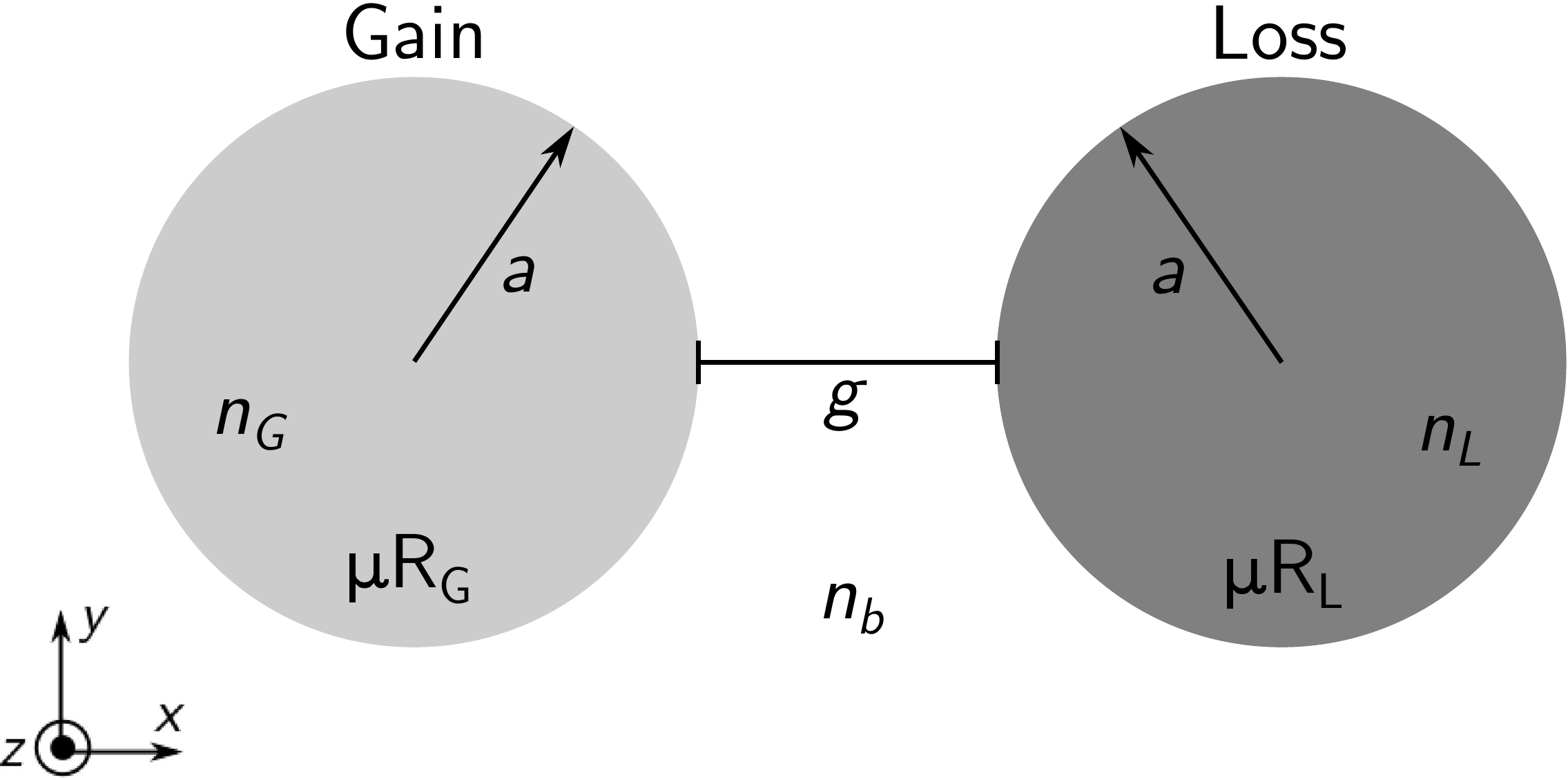}
	\end{overpic}
	\centering
	\caption[Schematic of \PT-symmetric microresonators.]{{Schematic of \PT-symmetric resonators. Microresonators with gain and loss are denoted by $\mu R_G$ and $\mu R_L$, respectively.}}
	\label{fig:10_illus}
\end{figure}

Having extended the TLM model for dispersive gain/loss material in 2D in the previous section, this section will develop the Boundary Integral Equation (BIE) \index{Boundary Integral Equation (BIE)} model for the \PT-coupled resonantor system. {Figure \ref{fig:10_illus} presents a schematic diagram of the system studied. It is comprised of  two microresonators,  each of radius $a$ and surrounded by air, that are separated by a gap $g$. The microresonators have complex refractive indices $n_G$ and $n_L$ respectively, that are chosen to satisfy the \PT-symmetric refractive index condition \index{\PT-symmetric refractive index} $n_G = n^*_L$, where $*$ denotes complex conjugate, $n = (n' + jn'')$, and $n'$ and $n''$ represent the real and imaginary parts of the refractive index.} The BIE model is developed for the complex refractive index which will be modelled by using the dispersive gain/loss model Eq. (\ref{eq:gainloss}). 

\subsection[Analysis of Inter-Resonator Coupling in the Frequency Domain]{{Inter-Resonator Coupling Model by Boundary Integral Equation} }
\label{sec:bieform}
{This section presents a Boundary Integral Equation (BIE) formulation to model the coupling between resonators. The BIE formulation is suited to a perturbative approximation of the coupling strength in the weak coupling limit, but also provides an efficient platform for exact calculation when coupling is strong. The approach pursued here is based on \cite{Creagh2001} which it was applied to describe coupling between fully bound states in coupled resonators and optical fibres. Here, the BIE is developed to allow for radiation losses similar to \cite{Boriskina2007a,Boriskina2006,Boriskina2007,Smotrova2006a,Smotrova2008}.}


The coupled \PT-microresonator \index{\PT-coupled resonator} depicted in Fig. \ref{fig:10_illus} is considered. For consistency the subscripts ``$G$" and ``$L$" are used for variables associated with the gain and lossy resonators respectively. Both resonators have  uniform refractive index, with the electric field polarized along the resonator axis. 

It is known that the electric field $\mathcal{E}_z$ takes the form $\psi_L=\frac{J_m(n_Lk_0r)}{J_m(n_Lk_0a)}\e^{jm\theta}$ inside the isolated lossy resonator and its normal derivative on the boundary of the resonator can be written as
\begin{align}\label{useF}
a\dydxv{\psi_L}{n} =  F^{L}_m \psi_L 
\end{align}
where 
\begin{align}\label{defF}
F^{L}_m = (n_{L}k_0a)\frac{J_m'(n_{L}k_0a)}{J_m(n_{L}k_0a)} 
\end{align}
where, $k_0$ is the free-space wave number and $\psi_G$ and $F^G_m$  are defined similarly for the gain resonator. It is emphasised that for notation simplicity, $\psi_L$ has been adopted to denotes the electric field $\mathcal{E}_z$ on the lossy resonator and $\psi_G$ for the electric field $\mathcal{E}_z$ on the gain resonator. The treatment of coupling presented in the remainder of this section can be used for other circularly-symmetric resonators such as those with graded refractive index or with different boundary conditions, as long as an appropriately modified $F^L_m$ is substituted in (\ref{useF}). 

\subsection{Graf's Addition Theorem}

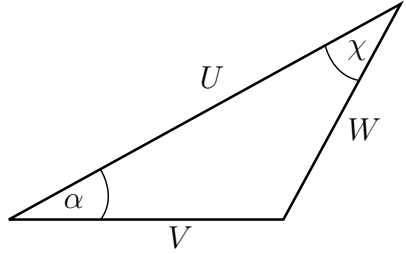
\begin{figure}[h]
	\begin{overpic}[width=0.45\textwidth,tics=5]{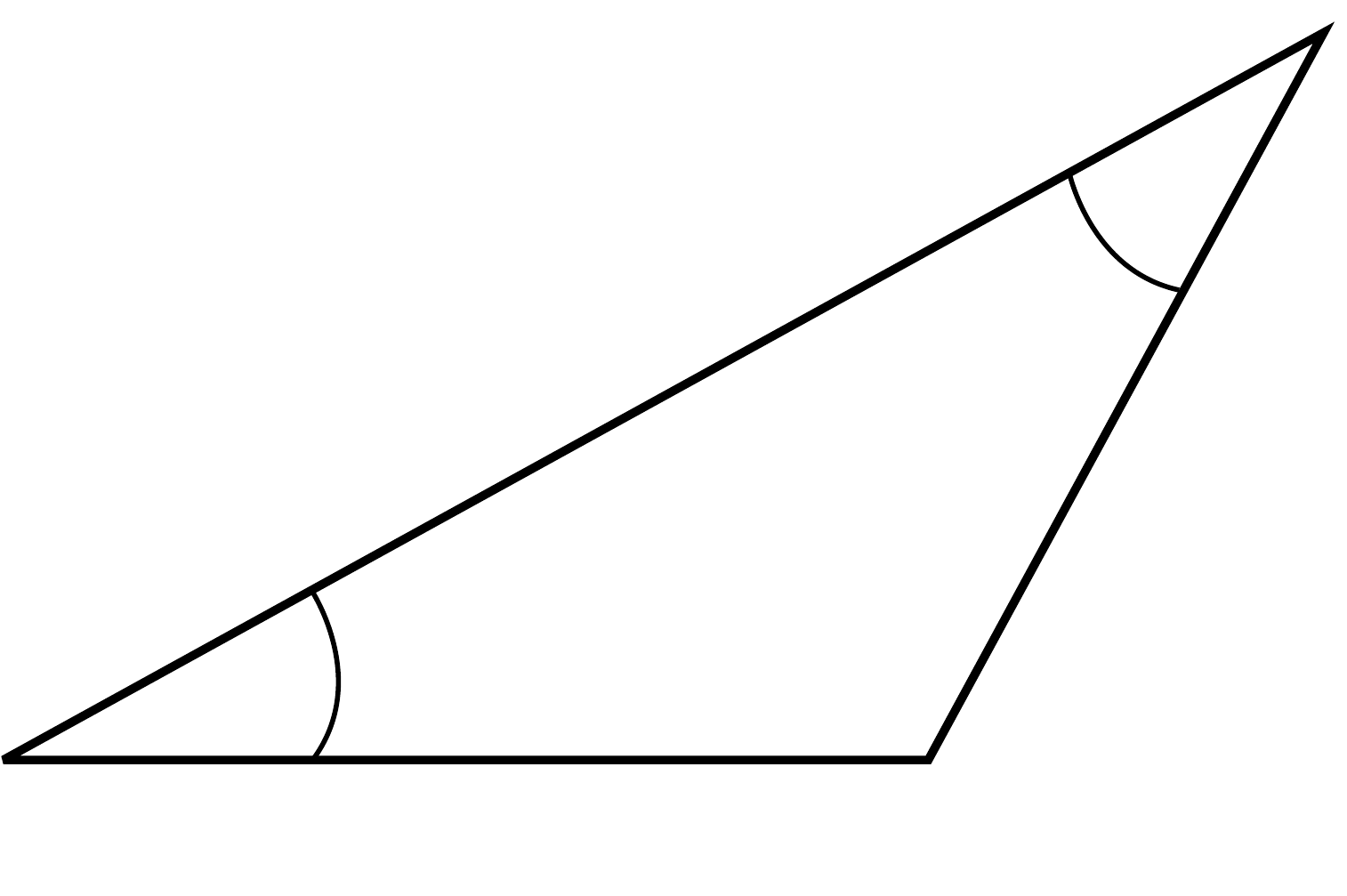}
		\put(14,13){\large$\alpha$}   \put(85,51.5){\large$\chi$}
		\put(48,43){\large$U$}  \put(40,3){\large$V$}  \put(85,30){\large$W$}
	\end{overpic}
	\centering
	\caption[Graf's addition theorem triangle.]{{Graf's addition theorem triangle.}}
	\label{fig:10_grafillus}
\end{figure}

As a prelude to the BIE model, the present subsection will overview Graf's addition theory \index{Graf's addition theory}. Graf's addition theory allows us to displace one cylindrical system of coordinates into another using Bessel function expansion \index{Bessel function expansion}. Consider $\mathscr{F}$, which can be any function from the Bessel function family $J$, $Y$, $H^\text{(1)}$, $H^\text{(2)}$ or indeed any linear combination of them. The following relation is valid \cite{Abramowitz1972}, 
\begin{align}
\mathscr{F}_m(W) e^{jn\chi} = \sum^{\infty}_{n=-\infty} \mathscr{F}_{m+n}(U) J_n(V) \e^{jn\alpha}, \quad \text{with} \quad U>V
\end{align}     
where $U$, $V$ and $W$ are real numbers defining distances. As such they can be interpreted as the edges of a triangle\cite{Abramowitz1972} as illustrated in Fig. \ref{fig:10_grafillus}.    

\subsection[Exact Solution Using Boundary-Integral Representation]{Exact Solution Using Boundary-Integral Representation }

In this subsection coupling between two dielectric circular resonators is studied using the Boundary Intergal Equation (BIE) \index{Boundary Integral Equation (BIE)} method. Expanding, the solution on each resonator boundary as a Fourier series,
\begin{align}
\psi_G = \sum_m \varphi_m^G \e^{jm\theta_G}
\qquad\mbox{and}\qquad
\psi_L = \sum_m \varphi_m^L \e^{jm\theta_L} 
\end{align}
in the polar angles $\theta_G$ and $\theta_L$ centred respectively on the gain and lossy resonators, running in opposite senses in each resonator and zeroed on the line joining the two centres. The corresponding normal derivatives at each boundary can be written as:
\begin{align}
\dydxv{\psi_G}{n} = \sum_m \frac{1}{a} F_m^G \varphi_m^G \e^{jm\theta_G}
\qquad\mbox{and}\qquad
\dydxv{\psi_L}{n} = \sum_m \frac{1}{a} F_m^L \varphi_m^L \e^{jm\theta_L} 
\end{align}

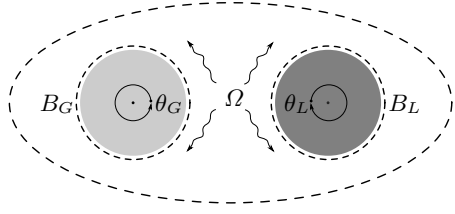
\begin{figure}[t]
	\begin{overpic}[width=0.5\textwidth,tics=5]{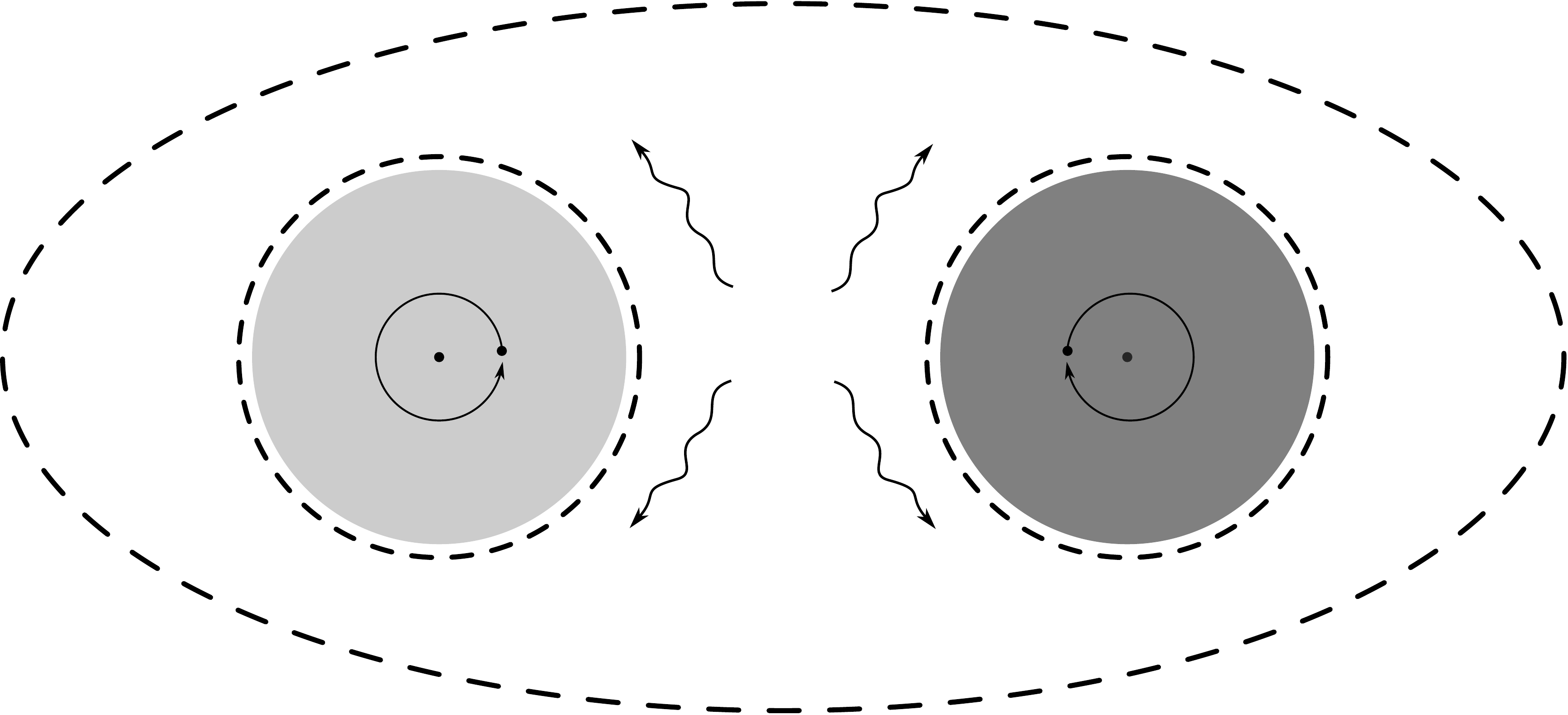}
		\put(48.5,22){$\Omega$} 
		\put(33,21){$\theta_G$} \put(62,21){$\theta_L$}
		\put(7,21){$B_G$} \put(85.5,21){$B_L$}
	\end{overpic}
	\centering
	\caption[Integration region $\Omega$ around the system of resonators]{{Integration region $\Omega$ around the system of resonators.} }
	\label{fig:10_illusproblemreg}
\end{figure}

An exact boundary integral representation of the coupled problem is conveniently achieved by applying Green's identities \index{Green's identities} to a region $\Omega$ which excludes the resonators, along an infinitesimally small layer surrounding them (so that the boundaries $B_G$ and $B_L$ of the resonators themselves lie just outside $\Omega$), see Fig. \ref{fig:10_illusproblemreg}. In $\Omega$, it is assumed that the refractive index $n_0=1$, such that the free-space Green's function \index{free-space Green's function} is \cite{Morse1953},
\begin{align}
G_0(\x,\x') = -\frac{j}{4} H_0 (k_0|\x-\x'|) 
\end{align}
where $\posx$ and $\posx'$ are the observation and the source points and $H_0(z) = J_0(z)-jY_0(z)$ denotes the Hankel function of the second kind (and the solution is assumed to have time dependence $\e^{j\omega t}$). Then, applying Green's identities to the region $\Omega$, and assuming radiating boundary conditions at infinity, leads to the equation,
\begin{align}
0 = \int_{B_G+B_L}
\left(G_0(\x,\x')\dydxv{\psi(\x')}{n'}-
\dydxv{G_0(\x,\x')}{n'}\psi(\x')\right)\d s'
\label{Green}
\end{align}
when $\x$ lies on either $B_L$ or $B_G$ (and therefore just outside of $\Omega$).

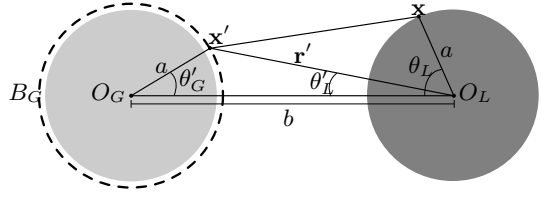
\begin{figure}[t]
	\begin{overpic}[width=0.6\textwidth,tics=5]{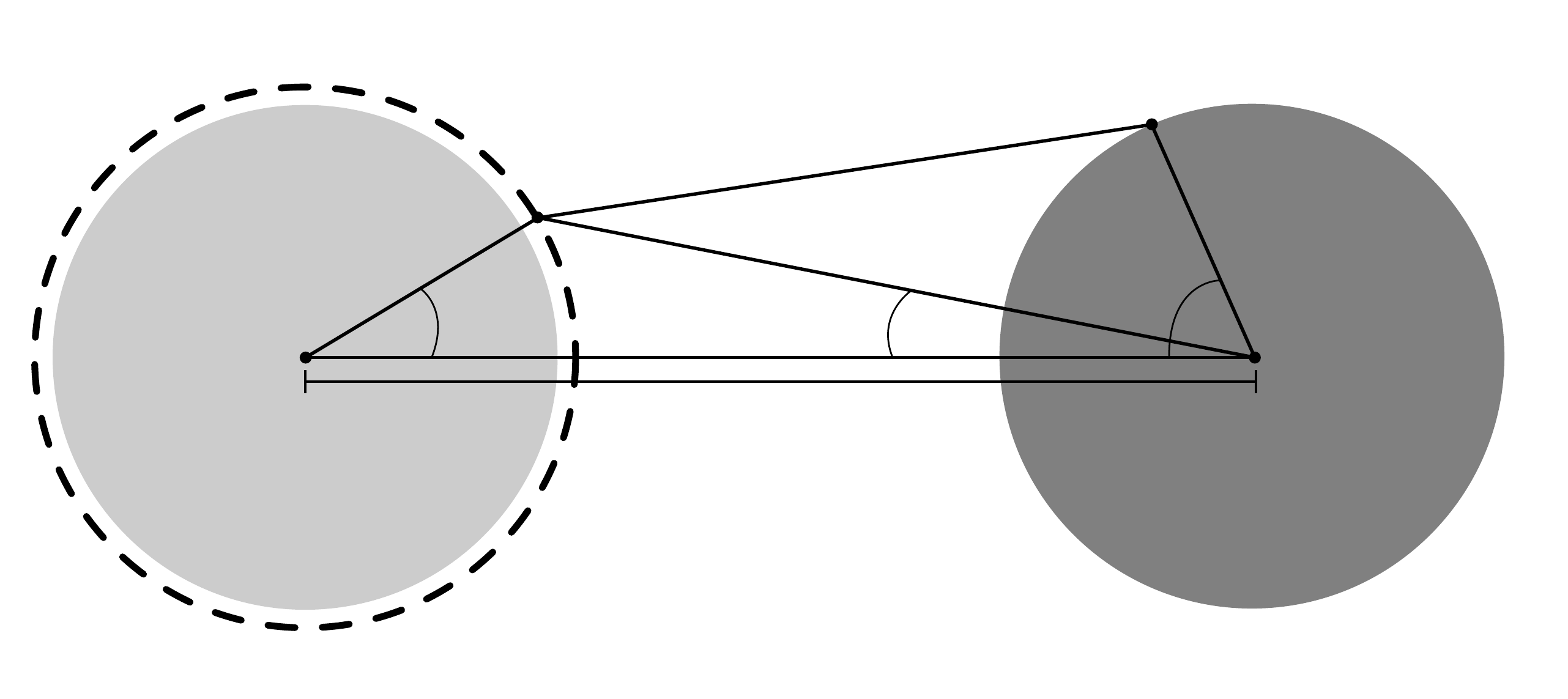}
		\put(48,16){$b$} \put(24,26){$a$} \put(77.5,29){$a$}
		\put(34,31.5){$\x'$} \put(72,37){$\x$}  \put(50,28){$\mathbf{r}'$}
		\put(-3.5,21){$B_G$}
		\put(12,21){$O_G$}  \put(81,21){$O_L$}
		\put(28.5,23.5){$\theta'_G$}  \put(71.5,26){$\theta_L$} \put(53,23){$\theta'_L$}
	\end{overpic}
	\centering
	\caption[Expansion of the free-space Green's function between the 2-coupled resonators by the Graf's addition theorem.]{{Expansion of the free-space Green's function between the two-coupled resonators by the Graf's addition theorem. Cross-contribution from the gain resonator to the lossy resonator.} }
	\label{fig:10_illusexpgreen}
\end{figure}

Using Graf's addition theorem \cite{Abramowitz1972}, the Green's function $G_0(\x,\x')$ is expanded analogously in polar coordinates on each boundary. First with respect to the triangle $\x'$ $O_L$ $\x$ (see Fig. \ref{fig:10_illusexpgreen}), it can be shown that,
\begin{align}
\begin{aligned}
H_0 (k_0|\x-\x'|) =\sum_\ell \left(H_\ell(k_0 \mathbf{r}')\e^{-j\ell\theta'_L}\right) J_\ell(k_0a) \e^{j\ell\theta_L} 
\end{aligned} \label{eq:greensexpand1}
\end{align}
Expanding the term in the bracket in Eq. (\ref{eq:greensexpand1}) with respect to triangle $O_G$ $\x'$ $O_L$ (see Fig. \ref{fig:10_illusexpgreen}), yields
\begin{align}
H_\ell(k_0 \mathbf{r}')\e^{-j\ell\theta'_L} = \sum_{\ell'}  H_{\ell+\ell'} (2bk_0) J_{\ell'}(k_0a) \e^{-j\ell'\theta'_G} 
\label{eq:greensexpand2}
\end{align}
Substituting Eq. (\ref{eq:greensexpand2}) into Eq. (\ref{eq:greensexpand1}), the Green's function can be expressed as, 
\begin{align}
G_0(\x,\x') = -\frac{j}{4} \sum_\ell \sum_{\ell'}  H_{\ell+\ell'} (2bk_0) J_{\ell'}(k_0a)  J_\ell(k_0a) \e^{j\ell\theta_L-j\ell'\theta'_G} 
\end{align}
with the corresponding normal derivatives of the Green's function, 
\begin{align}
\dfrac{\partial G_0(\x,\x')}{\partial n'} = -\frac{jk_0}{4} \sum_\ell \sum_{\ell'}  H_{\ell+\ell'} (2bk_0) J'_{\ell'}(k_0a)  J_\ell(k_0a) \e^{j\ell\theta_L-j\ell'\theta'_G} 
\end{align}

The Green's boundary integral on the lossy resonator due to the presence of gain resonator is
\begin{align}
\int_{B_G}
\left(G_0(\x,\x')\dydxv{\psi(\x')}{n'}-
\dydxv{G_0(\x,\x')}{n'}\psi(\x')\right)\d s'  
\label{eq:green1}
\end{align}
where the first term is calculated by, 
\begin{align}
\begin{aligned}
&\int_{B_G} G_0(\x,\x')\dydxv{\psi(\x')}{n'} \d s' \\
&= -\frac{j}{4a} \sum_{m \ell \ell'} \varphi_m^G F_m^G  H_{\ell+\ell'} (2bk_0) J_{\ell'}(k_0a)  J_\ell(k_0a)  \e^{j\ell\theta_L} 
\oint  \e^{j(m-\ell')\theta'_G} d\theta'_G 
\end{aligned}
\label{eq:greensfirstterm1}
\end{align}
Due to the orthogonality of the trigonometric function, equation (\ref{eq:greensfirstterm1}) can be simplified to, 
\begin{align}
\int_{B_G} G_0(\x,\x')\dydxv{\psi(\x')}{n'} \d s' = -j\frac{\pi}{2a} \sum_{m \ell} \varphi_m^G F_m^G  H_{\ell+m} (2bk_0) J_{m}(k_0a)  J_\ell(k_0a)  \e^{j\ell\theta_L}  
\label{eq:greensfirstterm2}
\end{align}
The second term of Eq. (\ref{eq:green1}) is calculated as, 
\begin{align}
\begin{aligned}
&\int_{B_G} \dydxv{G_0(\x,\x')}{n'}\psi(\x')\d s' \\
&= -\frac{jk_0}{4} \sum_{m \ell \ell'} \varphi_m^G  H_{\ell+\ell'} (2bk_0) J'_{\ell'}(k_0a)  J_\ell(k_0a) \e^{j\ell\theta_L} \oint   
\e^{j(m-\ell')\theta'_G} d\theta'_G 
\end{aligned}
\end{align}
which due to the orthogonality property of the trigonometric function can be simplified to, 
\begin{align}
\int_{B_G} \dydxv{G_0(\x,\x')}{n'}\psi(\x')\d s' = -j\frac{\pi k_0}{2} \sum_{m \ell} \varphi_m^G  H_{\ell+m} (2bk_0) J'_{m}(k_0a)  J_\ell(k_0a) \e^{j\ell\theta_L}  
\end{align}
The Green's boundary integral for the lossy resonator is now,
\begin{align}
\begin{aligned}
&\int_{B_G}
\left(G_0(\x,\x')\dydxv{\psi(\x')}{n'}-
\dydxv{G_0(\x,\x')}{n'}\psi(\x')\right)\d s' \\
&=  
-j\frac{\pi }{2a} \sum_{m \ell} \varphi_m^G  H_{\ell+m}(2bk_0) J_\ell(k_0a) \e^{j\ell\theta_L} 
\left[ F_m^G J_{m}(k_0a) - k_0a J'_{m}(k_0a) \right]   \\ 
\end{aligned}
\label{eq:greencross1}
\end{align}
Likewise, the Green's boundary integral on the gain resonator due to the presence of the lossy resonator, 
\begin{align}
\begin{aligned}
&\int_{B_L}
\left(G_0(\x,\x')\dydxv{\psi(\x')}{n'}-
\dydxv{G_0(\x,\x')}{n'}\psi(\x')\right)\d s' \\
&=  
-j\frac{\pi }{2a} \sum_{m \ell} \varphi_m^L  H_{\ell+m}(2bk_0) J_\ell(k_0a) \e^{j\ell\theta_G} 
\left[ F_m^L J_{m}(k_0a) - k_0a J'_{m}(k_0a) \right]  \\ 
\end{aligned}
\label{eq:greencross2}
\end{align}
As such Eq. (\ref{eq:greencross1}) describes the contribution of the gain resonator to the lossy resonator while Eq. (\ref{eq:greencross2}) describes the contribution of the lossy resonator to the gain resonator. 

\begin{figure}[t]
	\begin{overpic}[width=0.5\textwidth,tics=5]{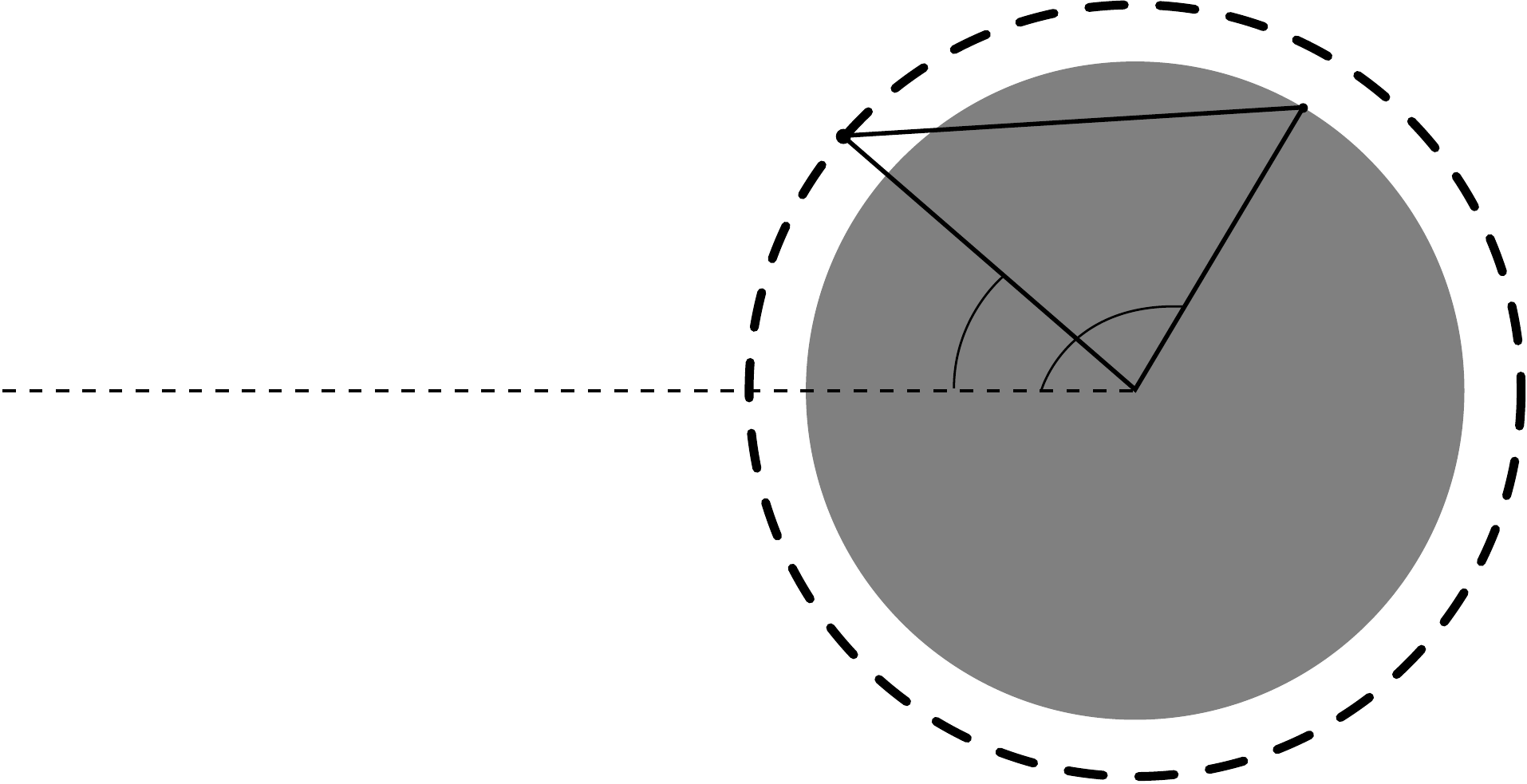}
		\put(53,44){$\x'$} \put(86,43){$\x$}  \put(65,35){$\mathbf{r}'$} \put(81,34){$a$}
		\put(75,21){$O_L$} \put(100,21){$B_L$}
		\put(71.5,32.5){$\theta_L$} \put(58,29){$\theta'_L$}
	\end{overpic}
	\centering
	\caption[Expansion of the self-contribution Green's function by Graf's addition theorem.]{{Expansion of the self-contribution Green's function by Graf's addition theorem.} }
	\label{fig:10_illusselfcont}
\end{figure}
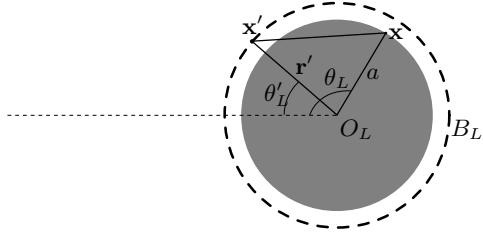

Following the cross-contribution to Green's integral, the self-contributions to Green's integral can be calculated. First consider only the lossy resonator as depicted in Fig. \ref{fig:10_illusselfcont}. The self-contribution of the lossy resonator can be calculated as,   
\begin{align}
\int_{B_L}
\left(G_0(\x,\x')\dydxv{\psi(\x')}{n'}-
\dydxv{G_0(\x,\x')}{n'}\psi(\x')\right)\d s'  
\label{eq:greenseq1}
\end{align}
As before, first expand the free-space Green's function \index{free-space Green's function} with respect to the triangle $\x'$ $O_L$ $\x$. By using the Graf's theorem \index{Graf's addition theory}, the Hankel function can be expanded as, 
\begin{align}
H_0 (k_0|\x-\x'|) &= \sum_\ell H_\ell(k_0 \mathbf{r}') J_\ell(k_0a) \e^{j\ell(\theta_L-\theta'_L)}  
\label{eq:greensexpand3}
\end{align}
As such the Green's function and its derivative at the boundary are given by
\begin{align}
G_0(\x,\x') &= -\frac{j}{4} \sum_\ell H_\ell(k_0 \mathbf{r}') J_\ell(k_0a) \e^{j\ell(\theta_L-\theta'_L)}   \\
\dfrac{\partial G_0(\x,\x')}{\partial n'} &= -j\frac{k_0}{4} \sum_\ell H'_\ell(k_0 a) J_\ell(k_0a)  \e^{j\ell(\theta_L-\theta'_L)}
\end{align}
Integrating the first term in Eq. (\ref{eq:greenseq1}) gives, 
\begin{align}
\begin{aligned}
\int_{B_L}&
G_0(\x,\x')\dydxv{\psi(\x')}{n'}
\d s'\\ 
&=-\frac{j}{4a} \sum_{m\ell} \varphi_m^L  F_m^L  H_\ell(k_0 a) J_\ell(k_0a) \e^{j\ell\theta_L} \oint\e^{j(m-\ell)\theta'_L} d\theta'_L \\
&=-j\frac{\pi}{2a} \sum_{m} \varphi_m^L  F_m^L  H_m(k_0 a) J_m(k_0a) \e^{jm\theta_L}  
\end{aligned}\label{eq:greenseq2}
\end{align}
and integrating the second term results in,
\begin{align}
\begin{aligned}
&\int_{B_L} \dydxv{G_0(\x,\x')}{n'}\psi(\x')\d s' \\
&=  -j\frac{k_0}{4} \sum_{m\ell} \varphi_m^L  H'_\ell(k_0 a) J_\ell(k_0a) \e^{j\ell\theta_L} \oint \e^{j(m-\ell)\theta'_L} d\theta'_L \\
&=  -j\frac{\pi k_0}{2} \sum_{m} \varphi_m^L  H'_m(k_0 a) J_m(k_0a) \e^{jm\theta_L}  
\end{aligned} 
\label{eq:greenseq3}
\end{align}
Hence the Green's integral due to the self-contribution of the lossy resonator is, 
\begin{align}
\begin{aligned}
&\int_{B_L}
\left(G_0(\x,\x')\dydxv{\psi(\x')}{n'}-
\dydxv{G_0(\x,\x')}{n'}\psi(\x')\right)\d s' \\
&= j \frac{\pi}{2a} \sum_{m} \varphi_m^L  J_m(k_0a)  \e^{jm\theta_L}  \left[ k_0a H'_m(k_0 a)    -  F_m^L  H_m(k_0 a) \right]       \\
\end{aligned} 
\label{eq:greenseq4}
\end{align} 
Likewise, the self-contribution of the gain resonator can be obtained as, 
\begin{align}
\begin{aligned}
&\int_{B_G}
\left(G_0(\x,\x')\dydxv{\psi(\x')}{n'}-
\dydxv{G_0(\x,\x')}{n'}\psi(\x')\right)\d s' \\
&= j \frac{\pi}{2a} \sum_{m} \varphi_m^G  J_m(k_0a)  \e^{jm\theta_G}  \left[ k_0a H'_m(k_0 a)    -  F_m^G  H_m(k_0 a) \right]       \\
\end{aligned} 
\label{eq:greenseq5}
\end{align} 
Summing the self-contribution and cross-contribution for each resonator and substituting it into (\ref{Green}), it can be shown that the Green's boundary integral for the gain resonator, 
\begin{align}
\begin{aligned}
&\sum_{m} J_m(k_0a)    H_m(k_0 a) \left[ F_m^G - k_0a \frac{H'_m(k_0 a)}{H_m(k_0 a)}   \right] \varphi_m^G  \\
+&\sum_{m \ell} J_{m}(k_0a)  H_{\ell+m}(2bk_0) J_\ell(k_0a) 
\left[ F_m^L  - k_0a \frac{J'_{m}(k_0a)}{J_{m}(k_0a)} \right]  \varphi_m^L
= 0        \\
\end{aligned} 
\label{eq:greenseq6}
\end{align} 
and for the lossy resonator,
\begin{align}
\begin{aligned}
&\sum_{m} J_m(k_0a)    H_m(k_0 a) \left[ F_m^L - k_0a \frac{H'_m(k_0 a)}{H_m(k_0 a)}   \right] \varphi_m^L  \\
+&\sum_{m \ell} J_{m}(k_0a)  H_{\ell+m}(2bk_0) J_\ell(k_0a) 
\left[ F_m^G  - k_0a \frac{J'_{m}(k_0a)}{J_{m}(k_0a)} \right]  \varphi_m^G
= 0        \\
\end{aligned} 
\label{eq:greenseq7}
\end{align} 
Equations (\ref{eq:greenseq6}) and Eq. (\ref{eq:greenseq7}) can also be expressed in matrix form as,  
\begin{align}\label{sys1} 
\begin{split}
D^G\varphi^G+ C^{GL} \varphi^L&= 0\\ 
C^{LG}\varphi^G+ D^L \varphi^L&= 0 
\end{split}
\end{align} 
where, 
\begin{align}
\varphi^G = \left(\begin{array}{c}\vdots\\\varphi_m^G\\\varphi_{m+1}^G\\\vdots\end{array}\right)
\qquad\mbox{and}\qquad
\varphi^L = \left(\begin{array}{c}\vdots\\\varphi_m^L\\\varphi_{m+1}^L\\\vdots\end{array}\right)
\end{align}   
are Fourier representations of the solution on the boundaries of the gain and lossy resonators respectively. The matrices $D^G$ and $D^L$ are diagonal with entries 
\begin{align}
D^{G,L}_{mm} =
J_m(u)H_m(u)\left(F_m^{G,L}-\frac{uH_m'(u)}{H_m(u)}\right),
\qquad\mbox{where $u = k_0a$} 
\end{align}
and provide the solutions for the isolated resonators.  The matrices
$C^{GL}$ and $C^{LG}$ describe coupling between the resonators. The matrix
$C^{GL}$ has entries of the form
\begin{align}
C_{lm}^{GL} = J_l(u)H_{l+m}(w)
J_m(u)
\left(F_m^{L}-\frac{u J_m'(u)}{J_m(u)}\right) 
\end{align}
where $u=k_0a$, $w = k_0b$ and $b$  is the centre-centre distance between the gain and lossy resonators as indicated in Fig. \ref{fig:10_illus}. The matrix $C^{LG}$ is defined manner by swapping the labels $G$ and $L$. 

The system (\ref{sys1}) can be presented more symmetrically by using the
scaled Fourier coefficients
\begin{align}
\tilde{\varphi}^L_m =
J_m(u)
\left(F_m^L-\frac{uJ_m'(u)}{J_m(u)}\right)
\varphi_m^L
\end{align}
(along with an analogous definition of $\tilde{\varphi}_m^G$). Then
(\ref{sys1}) can be rewritten as
\begin{align}
\begin{split}
\tilde{D}^G\tilde{\varphi}^G+ \tilde{C} \tilde{\varphi}^L&= 0\\ 
\tilde{C}\tilde{\varphi}^G+ \tilde{D}^L \tilde{\varphi}^L&= 0 
\end{split}
\label{sys2} 
\end{align} 
where the diagonal matrices $\tilde{D}^{G,L}$ have entries
\begin{align}
\tilde{D}^{G,L}_{mm} =
-j \frac{H_m(u)F_m^{G,L}-uH_m'(u)}{J_m(u)F_m^{G,L}-uJ_m'(u)},
\qquad\mbox{where $u = k_0a$} 
\end{align}
and the matrix $\tilde{C}$, with entries
\begin{align}
\tilde{C}_{lm} =-j H_{l+m}(w) 
\end{align}
couples solutions in both directions. 

A factor of $-j$ is included in these equations to highlight an approximate \PT-symmetry \index{\PT-symmetry} that occurs when $n_G=n_L^*$. In the limit of high-$Q$ (low loss) whispering gallery resonances \index{whispering gallery resonances}, the following approximations hold,
\begin{align}
jH_m(u)\simeq Y_m(u)\qquad\mbox{and}\qquad
jH_{l+m}(u)\simeq Y_{l+m}(u) 
\end{align}
and the matrices in (\ref{sys2}) satisfy the conditions
\begin{align}
\left(\tilde{D}^L\right)^* \simeq \tilde{D}^G 
\qquad\mbox{and}\qquad
\tilde{C}^* \simeq \tilde{C} 
\end{align}
which are a manifestation of \PT-symmetry \index{\PT-symmetry} of the system as a
whole: deviation from  these conditions is a consequence of the radiation losses.

\subsection{Weak-Coupling Perturbation Approximation}
We can exploit (\ref{sys2}) to form an
efficient numerical method for determining the resonances of the
coupled system to arbitrary accuracy. {It was observed that in practice truncation of the system to relatively few modes was sufficient to describe the full solution once the gap $g = b - 2a$ between the resonators was wavelength-sized or larger.} 

For very weak coupling, an effective perturbative approximation can be achieved by restricting our consideration to a single mode in each resonator. We consider in particular the case of near left-right symmetry in which 
\begin{align}
n_G\approx n_L  
\end{align}
\PT-symmetry is achieved by further imposing $n_G=n_L^*$, but for
now the effects of dispersion are allowed by assuming that this is not
the case. The full solution is built around modes for which
\begin{align}
\psi_{\pm} \approx \psi_G \pm \psi_L 
\end{align}
where $\psi_G$ and $\psi_L$ are the solutions of the isolated
resonators described at the beginning of this section. A single value of $m$ is used for both $\psi_G$ and $\psi_L$ and in 
particular the global mode is approximated using a chiral state in 
which the wave circulates in opposite senses in each resonator.
That is, the coupling between $m $ and $-m$ that occurs in
the exact solution is neglected.

Then a simple perturbative approximation is achieved by
truncating the full system of (\ref{sys2}) to the $2\times 2$ system
\begin{align}
M
\left(
\begin{array}{c}\tilde{\varphi}^G_{mm}\\ \tilde{\varphi}^L_{mm}\end{array}
\right) = 0,
\qquad\mbox{where}\quad
M=\left(
\begin{array}{cc}\tilde{D}^G_{mm}&\tilde{C}_{mm}\\\tilde{C}_{mm}&\tilde{D}^L_{mm}\end{array}
\right) 
\label{eq:mateq}
\end{align}
Resonant frequencies of the coupled problem are then realised when
\begin{align}
0 = \det M = \tilde{D}^G_{mm}\tilde{D}^L_{mm}-\tilde{C}_{mm}^2 
\end{align}
In the general, dispersive and non-\PT-symmetric, {case the calculation is reduced to a semi-analytic solution in which the (complex) roots of the known $2\times 2$ determinant in Eq. (\ref{eq:mateq}) are required, and in} which the matrix elements depend on
frequency through both $k_0=\omega/c$ and $n=n(\omega)$. 

\subsection[Further Analytic Development of the Perturbative Solution]{\PT-Symmetric Threshold of Weakly-Coupled System}

To develop a perturbative expansion let, 
\begin{align}
D_{mm}^0 = \frac{1}{2}\left(\tilde{D}_{mm}^G+\tilde{D}_{mm}^L\right) 
\qquad \mbox{and} \qquad
D_{mm}^I = \frac{1}{2j}\left(\tilde{D}_{mm}^G-\tilde{D}_{mm}^L\right)
\end{align} 
(and note that in the high-$Q$-factor \PT-symmetric case,
$\tilde{D}^G\simeq(\tilde{D}^L)^*$, both $D_{mm}^0$ and $D_{mm}^I$ are
approximately real). It is assumed that both $D^I_{mm}$ and $C_{mm}$ are
small and comparable in magnitude. Expand the angular frequency
\begin{align}
\omega_{1,2} = \omega_0 \pm \frac{\Delta\omega_0}{2} + \cdots
\end{align}
about a real resonant angular frequency of an averaged isolated
resonator satisfying
\begin{align}
D^0_{mm}(\omega_0) = 0 
\end{align}
Then to first order of accuracy the coupled resonance condition becomes
\begin{align}
0=\det M = \Delta\omega_0^2 {D_{mm}^0}'(\omega_0)^2 +
D_{mm}^I(\omega_0)^2-\tilde{C}_{mm}(\omega_0)^2 + \cdots
\end{align}
from which the angular frequency shifts can be written as\cite{phang2015b,SendyPhang2016},
\begin{align}\label{getomega1}
\frac{\Delta\omega_0}{2} = \frac{\sqrt{\tilde{C}_{mm}(\omega_0)^2-D_{mm}^I(\omega_0)^2}}{{D_{mm}^0}'(\omega_0)} 
\end{align}
where ${D_{mm}^0}'(\omega)$ denotes a derivative of $D_{mm}^0(\omega)$
with respect to frequency.

The simple condition
\begin{align}
\tilde{C}_{mm}(\omega_0)^2=D_{mm}^I(\omega_0)^2 
\end{align}
is obtained for the threshold at which $\Delta\omega_0=0$ and the two resonant frequencies
of the coupled system coincide. In the \PT-symmetric case, where
$\tilde{C}_{mm}$ and $D_{mm}^I$ are approximately real (and whose
small 
imaginary parts represent corrections due to radiation 
losses), we therefore have a prediction for a real threshold frequency results. \index{threshold frequency}

\section{Symmetry breaking in \PT-Microresonator Couplers}
{In this section, we employ the Boundary Integral Equation (BIE) and the numerical Transmission-Line Modelling (TLM) method to investigate the impact of gain/loss material parameters, such as the dispersion parameter and the gain/loss parameter, on the spectra of the \PT-Microresonator Couplers and show the relation between the operating mode and the threshold behaviour of such structures. }

\begin{comment}
In this section, the impact of dispersion on the resonant frequencies and threshold behaviour of the \PT-symmetric microresonators \index{PT-coupled resonator} is analysed. Frequency mismatch between the resonant frequency of the microresonator and gain pump frequency is investigated for practical levels of dispersion and the practical implications of a slight unbalance between the gain and loss in the system are investigated.  The section concludes with an investigation of how coupling between resonators manifests itself in the time development of solutions.
\end{comment}

\subsection[Impact of Gain/Loss Material Parameters on Threshold Behaviour in the Frequency Domain]{Impact of Gain/Loss Material Parameters on Threshold Behaviour in the Frequency Domain} 

{For definiteness, the specifications of the \PT-microresonator coupler investigated in this section are now given: coupled microresonators weakly coupled via their evanescent fields separated by a distance $g = 0.24\text{ }\mu\mbox{m}$, and made of GaAs material of dielectric constant $\varepsilon_\infty =3.5$ \cite{Hagness1996,phang2015b} with radius $a = 0.54\text{ }\mu\mbox{m}$ (see Fig. \ref{fig:10_illus}). As an isolated passive microresonator, two closely spaced whispering-gallery modes exist, namely a low $Q$-factor \index{$Q$-factor} Whispering-Gallery Mode - WGM(7,2) and a high $Q$-factor mode WGM(10,1), whose resonant frequencies are respectively $f_0^{(7,2)}=341.59\mbox{ THz}$ and $f_0^{(10,1)}=336.85\mbox{ THz}$, with $Q$-factors $Q^{(7,2)}=2.73\times10^3$ and $Q^{(10,1)}=1.05\times10^7$ }

\begin{comment}
For all cases, weakly coupled microresonators are considered, the coupled resonators each have a dielectric constant $\varepsilon_\infty
=3.5$ \cite{Hagness1996,phang2015b} radius $a = 0.54\text{ }\mu\mbox{m}$ and are separated by distance $g = 0.24\text{ }\mu\mbox{m}$. Operation at two different whispering-gallery modes is analysed, namely a low $Q$-factor \index{$Q$-factor} mode (7,2) and a high $Q$-factor mode (10,1). The corresponding isolated resonator resonant frequencies are respectively $f_0^{(7,2)}=341.59\mbox{ THz}$ and $f_0^{(10,1)}=336.85\mbox{ THz}$, with $Q$-factors $Q^{(7,2)}=2.73\times10^3$ and $Q^{(10,1)}=1.05\times10^7$. 
\end{comment}

\begin{figure}[]
	\centering
	\begin{overpic}[width=0.8\textwidth,tics=5]{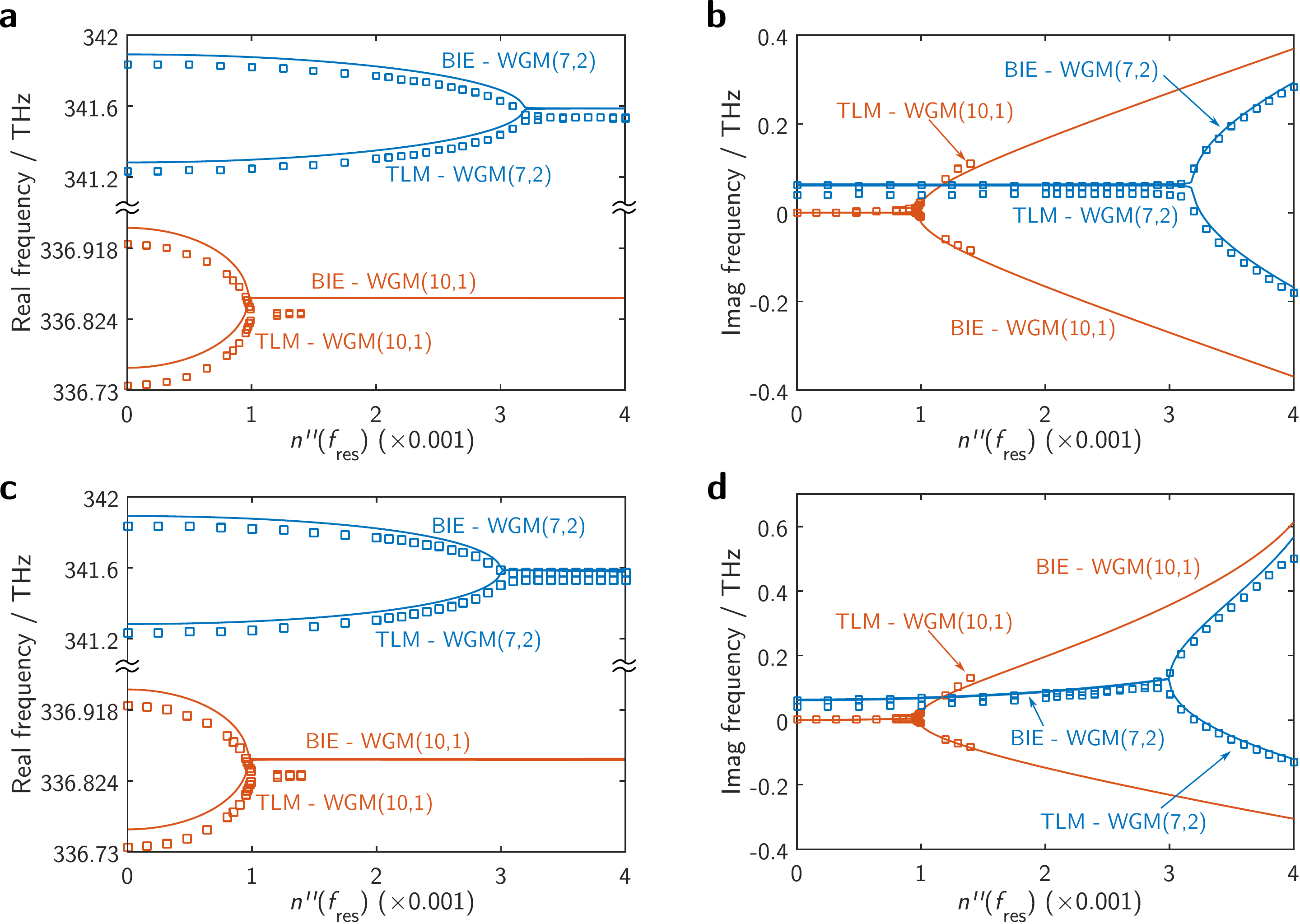}
	\end{overpic}
	\caption[Frequency bifurcation of \PT-coupled microresonator]{{Frequency bifurcation of \PT-coupled microresonator. Resonators have balanced gain/loss ($n''_G=-n''_L$). The plots shows the real and imaginary part of the resonant frequenices calculated by both the Boundary Integral Equation (BIE) and the numerical Transmission-Line Modelling (TLM) method. Results from the BIE are shown by solid line and from the TLM by discrete points. These are displayed as a function of gain/loss parameter calculated at the peak of pumping beam $n''(f_\text{res})$ for three different dispersion parameters, ({a,b}) $\omega_\sigma\tau=0$ and ({c,d}) $\omega_\sigma\tau=212$.}}
	\label{fig:10_spliting}
\end{figure}

{The impact of gain/loss parameter $n''(f_\text{res})$ on the eigenfrequencies of the \PT-microresonator coupler is presented in Fig. \ref{fig:10_spliting} for two different dispersion parameters. Figure \ref{fig:10_spliting}(a,b) presents the dispersionless case, $\omega_\sigma\tau=0$ and Fig. \ref{fig:10_spliting}(c,d) uses dispersion values typical of GaAs, $\omega_\sigma\tau=212$ \cite{Hagness1996}. In this figure, we considered the case of equal material gain and loss $n''_G=-n''_L$. Moreover, the gain and loss at the atomic transitional (angular) frequency $\omega_\sigma$ are assumed to be tuned to the resonant frequency of the isolated case, i.e. $\omega_\sigma=2\pi f_\text{res}$; for the case when $\omega_\sigma$ is \textit{not} in tune, i.e. $\omega_\sigma \ne 2\pi f_\text{res}$ the reader is referred to \cite{phang2015b}. }  

\begin{comment}
Figure \ref{fig:10_spliting} shows the real and imaginary parts of the eigenfrequencies $f_1$ and $f_2$ of the \PT-symmetric coupled microresonators with balanced gain and loss, $n''_G=-n''_L$, and is depicted as a function of the imaginary part of the refractive index $n''(f_\text{res})$ for both the low and high $Q$-factor modes. The gain and loss are assumed to be tuned to the resonant frequency of an isolated microresonator, i.e. $\omega_\sigma=2\pi f_\text{res}$. Three different levels of dispersion, controlled by the parameter $\tau$ are considered. These are $\omega_\sigma\tau=0$ corresponding to the case of no dispersion and $\omega_\sigma\tau=212$ taken from \cite{Hagness1996} to exemplify the case of a realistic dispersion parameter.
\end{comment}

{Figure \ref{fig:10_spliting} shows that the coupling to the other microresonator leads to the formation of super-modes each of which is centred at the resonant frequency of the isolated case. The increase of gain/loss in the system causes the super-modes to beat at a lower rate (i.e. the difference in the super-mode resonant frequencies decreases) which leads the super-modes to coalesce at the threshold point, i.e. $n''(f_\text{res})=0.0032$ and 0.001 for the low $Q$-factor WGM(7,2) and the high $Q$-factor WGM(10,1) respectively, see Fig. \ref{fig:10_spliting}(a and c). Operation with gain/loss above this threshold point leads to unstable operation, indicated by the splitting of the imaginary part of the eigenfrequencies (see Fig. \ref{fig:10_spliting}(b and d)). It is noted here that for a fixed separation distance $g$, the high-$Q$ factor WGM(10,1) mode has a lower threshold compared to the low-$Q$ factor WGM(7,2) mode regardless of the material dispersion parameter.} 

{The imaginary part of the eigenfrequency for the low $Q$-factor WGMs, depicted in Fig. \ref{fig:10_spliting}(b and d), shows significant positive values of the imaginary part before the threshold point, in comparison to the high $Q$-factor WMGs. This signifies the high intrinsic radiation losses. Visual inspection between Fig. \ref{fig:10_spliting}(b) and Fig. \ref{fig:10_spliting}(d) shows that dispersion modifies the imaginary part of the eigenfrequency in such a way that before the threshold point it is no longer constant, as in the dispersionless case, but skewed towards a higher value of overall loss. After the threshold point the imaginary part does not equally split to form complex conjugate \index{complex conjugates} eigenfrequencies, as in the dispersionless case, but is also skewed towards a higher value of overall loss.  }

Furthermore, Figs. \ref{fig:10_spliting}(a-d) compares the eigenfrequencies calculated by the Boundary Integral Equation (BIE) \index{Boundary Integral Equation (BIE)}, i.e. zeros of the linear problem (\ref{sys2}), and the time-domain 2D-TLM method described in Section \ref{sec:2dtlm}. The TLM method simulates the same problem as the BIE counterpart except that the TLM model introduces spatial discretisation for which in these calculations $\Delta \ell = 2.5 \times 10^{-3}\text{ }\mu\text{m}$ is used\footnote{This discretisation parameter is equivalent with $\lambda_\text{sim}/100$, where $\lambda_\text{sim}$ is the maximum simulation bandwidth in material, i.e. $\lambda_\text{sim}=0.875\text{ }\mu\text{m}/3.5$.}. The TLM model uses an electric dipole excitation with a Gaussian profile modulated at the resonant frequency of the isolated resonator $f_\text{res}$ with FWHM of 250 fs to provide a narrow bandwidth source for a total simulation time of 3 ps. The complex eigenfrequencies are extracted by using the Harmonic inversion method \index{Harmonic inversion} \cite{grossman97,Mandelshtam1998,Johnson2015,Vukovic2014}; for these calculations the freely available Harminv package \cite{Johnson2015} was used. Details of the harmonic inversion by filter diagonalisation method are not described in this chapter and reader is referred to \cite{grossman97,Mandelshtam1998,Johnson2015}; the software package used in this work is freely available to download\footnote{{http://ab-initio.mit.edu/wiki/index.php/Harminv}}.     

By comparing the eigenfrequencies calculated by the BIE and the TLM method (discrete bullet points), it can be seen from  Figs. \ref{fig:10_spliting}(a,c) that the real part of the eigenfrequencies calculated by the TLM method are shifted to the lower frequencies (red-shifting) which occurs due to numerical dispersion and stair-casing approximation. It is noted that a similar red-shifting error \index{red-shifting error} was also observed during the investigation of \PT-Bragg grating \index{\PT-Bragg grating} in the previous chapter using the TLM method. This error can be minimised by reducing the mesh discretisation length with the cost of longer CPU simulation time. Nevertheless, Figs. \ref{fig:10_spliting}(a,c) show that both the TLM and the BIE calculations predict and follow the same threshold behaviours \index{threshold behaviour}. A more detail temporal analysis using the TLM model will be discussed in the next section.    

\begin{comment}
\begin{figure}[t]
	\begin{overpic}[width=0.8\textwidth,tics=5]{figures/10_mismatch}
	\end{overpic}
	\centering
	\caption[Mismatching gain/loss material transitional frequency and the resonator resonant frequency.]{{Mismatching gain/loss material transitional frequency and the resonator resonant frequency. Frequency bifurcation of coupled microresonators with balanced gain and loss as function of gain/loss parameters $n''(\omega_\sigma)$, for two different atomic transitional frequencies $\omega_\sigma=2\pi(f_\text{res}+\delta)$ with $\delta =-0.1$ and $0.1\text{ THz}$.}}
	\label{fig:10_mismatch}
\end{figure}
\end{comment}

\begin{figure}[t]
	\begin{overpic}[width=0.85\textwidth,tics=5]{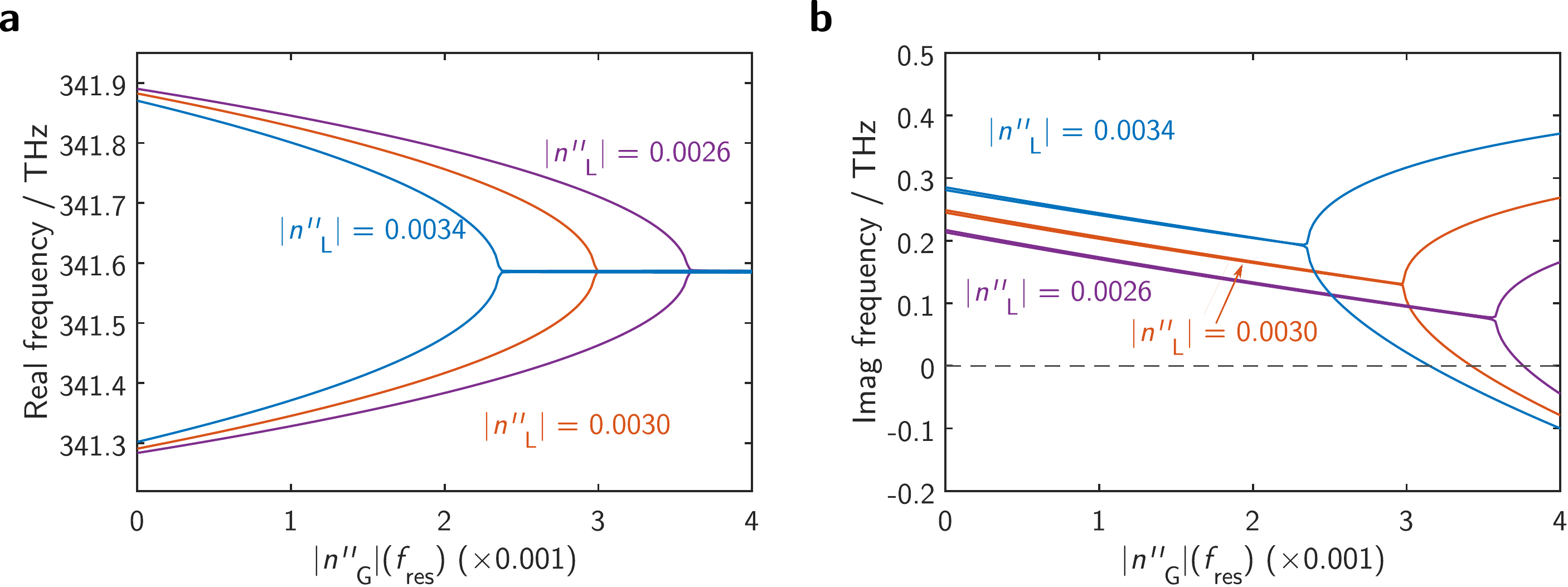}
	\end{overpic}
	\centering
	\caption[Complex eigenfrequency in a \PT-coupled microresonator system with variable gain and fixed loss]{{Complex eigenfrequency in a \PT-coupled microresonator system with variable gain and fixed loss, plotted as a function of gain parameter $|n''_G|$, dispersion parameter $\omega_\sigma\tau=212$\cite{Hagness1996} and shown for three different fixed loss value, i.e. $|n''_L|=0.0026$, $0.0030$, and $0.0034$.}}
	\label{fig:10_vargain}
\end{figure}

The key conclusion to be made from Fig. \ref{fig:10_spliting} is therefore that \PT-like threshold behaviour is observed in the cases of no dispersion and of high dispersion. While there is some skewness in the high-dispersion case, which amounts to a quantitative deviation from strict \PT-symmetry, there is an essential qualitative similarity to the dispersionless case in which there appears to be a sharp threshold. 

\begin{comment}
Having confirmed that realistic levels of dispersion preserve \PT-behaviour, Figure \ref{fig:10_mismatch} considers a practical scenario in which there is high dispersion $\omega_\sigma\tau=212$ and a frequency mismatch between the resonant frequency and the gain/loss atomic angular frequency. The material atomic frequency is defined to be $\omega_\sigma=2\pi(f_\text{res}+\delta)$, where $\delta$ is the mismatch parameter. The structure is operated with balanced gain and loss, i.e. $n''_G=-n''_L$ and two values are assumed for the frequency mismatch, namely $\delta=-0.1$ and 0.1 THz. Figure \ref{fig:10_mismatch}(a,b) shows the results for the low $Q$-factor mode (7,2) and Fig. \ref{fig:10_mismatch}(c,d) for the high-$Q$ factor mode (10,1). In both cases there is no sharp threshold point for the real parts of eigenfrequencies and the imaginary parts begin to diverge at low gain/loss values. Neither are the imaginary parts symmetrically placed about a branching value. This result confirms the fact that \PT-behaviour is preserved only when the angular transitional frequency \index{transitional frequency} of the dispersive gain/loss profile is aligned with the resonant frequency of the microresonators. If this is not the case, the frequency misalignment causes the coupled system to continue to beat after a threshold region.
\end{comment}

{In Figs. \ref{fig:10_spliting}, we have considered the impact on gain/loss parameter on the spectra of the \PT-microresonator coupler for the case when the material gain/loss is equal, i.e. $n''_G=-n''_L$. The case when the material gain is not equal is now considered. Figure \ref{fig:10_vargain}(a,b) shows the eigenfrequency of the \PT-microresonator coupler for three different values of loss namely $|n''_L|=0.0026$, 0.0030 and 0.0034, which correspond to values below, at, and above the threshold point of a \PT-symmetric structure \index{\PT-symmetric structure} with equal gain and loss respectively. The low $Q$-factor WGM (7,2) is considered with practical material dispersion parameters. }

{Figure \ref{fig:10_vargain} shows that there exists a threshold point even when the gain and loss parameter are not equal. More interestingly, this figure further shows that by increasing loss, the threshold point of the system, i.e. the level of gain needed to achieve lasing, decreases. This counter-intuitive principle of attaining lasing operation by increasing loss has been experimentally shown in \cite{Peng2014} in which they use a metal probe to increase loss in the lossy microresonator. }

\begin{comment}
Another practical scenario is considered in Fig. \ref{fig:10_vargain} where the gain and loss are not balanced, i.e.  $\mu\text{R}_L$ has a loss $|n''_L|$ while $\mu\text{R}_G$ has a gain $|n''_G|$. Figure \ref{fig:10_vargain}(a,b) shows the real and imaginary parts of the eigenfrequency for three different values of loss namely, $|n''_L|=0.0026$, 0.0030 and 0.0034 which correspond to values below, at, and above the threshold point of a \PT-symmetric structure \index{\PT-symmetric structure} with balanced gain and loss respectively. The low $Q$-factor \index{$Q$-factor} mode is considered with a practical dispersion parameter of $\omega_\sigma\tau=212$ as taken from \cite{Hagness1996}. Interestingly, it can now be observed that the \PT-threshold \index{\PT-threshold} point can also exist for structures with unbalanced gain/loss as shown by the plots for $|n''_L|=0.0026$ and $|n''_L|=0.0034$ in Fig. \ref{fig:10_vargain}. In the former case, the \PT-threshold is increased and in the latter case the \PT-threshold is decreased when compared with the \PT-threshold of the balance structure. Of special interest is the observation that increasing loss results in the reduction of the \PT-threshold which consequently reduces the levels of gain at which lasing occurs. This counter-intuitive principle of switching lasing on by increasing loss has been experimentally demonstrated in \cite{Peng2014} where a metal probe was used to enhance loss in the lossy microresonator. 
\end{comment}

\subsection[Real Time Operation of \PT-Coupled Microresonators]{Real Time Operation of \PT-Microresonators Coupler}

{In \cite{phang2015b}, the authors have studied the impact of material dispersion on the real-time operation of a \PT-microresonator coupler. In this subsection, we will summarise the important results of that study and present the temporal dynamics of a \PT-microresonator coupler in a practical scenario. The interested reader is referred to \cite{phang2015b}. }

{The real-time operation of the \PT-microresonator coupler is analysed in the time-domain employing the two-dimensional 2D Transmission-Line Modelling \index{Transmission-Line Modelling (TLM)} (2D-TLM) method which was described in detail in Section \ref{sec:2dtlm}. In all cases it is found that the TLM simulations agree with the frequency-domain calculations provided in the previous section, and in fact have been used to independently validate the BIE \index{Boundary Integral Equation (BIE)} analysis, comparisons with which are presented in Figs. \ref{fig:10_spliting}. The TLM simulations only considered the low $Q$-factor WGM(7,2) mode which was excited by a narrow bandwidth Gaussian dipole source tuned at the resonant frequency of the WGM(7,2).  }

{In summary, \cite{phang2015b} shows that material dispersion is essential in ensuring the stability of operation of the \PT-microresonator coupler. This is due to the fact that in the \textit{idealised} dispersionless case violates the Kramers-Kronig \index{Kramers-Kronig} relationship\cite{Zyablovsky2014,Phang2014d,phang2015b}. This violation causes the \PT-symmetric condition to be satisfied throughout the frequency spectrum, allowing the existence of an infinite number of threshold (exceptional) points which leads to multi-frequency lasing. Contrary to this, when practical material dispersion is considered, the \PT-symmetric condition is satisfied \textit{only} at a single frequency which is tuned to the atomic transitional frequency of the material gain/loss \cite{Phang2014d,phang2015b}.} 

\begin{figure}[h!]
	\begin{overpic}[width=0.87\textwidth,tics=5]{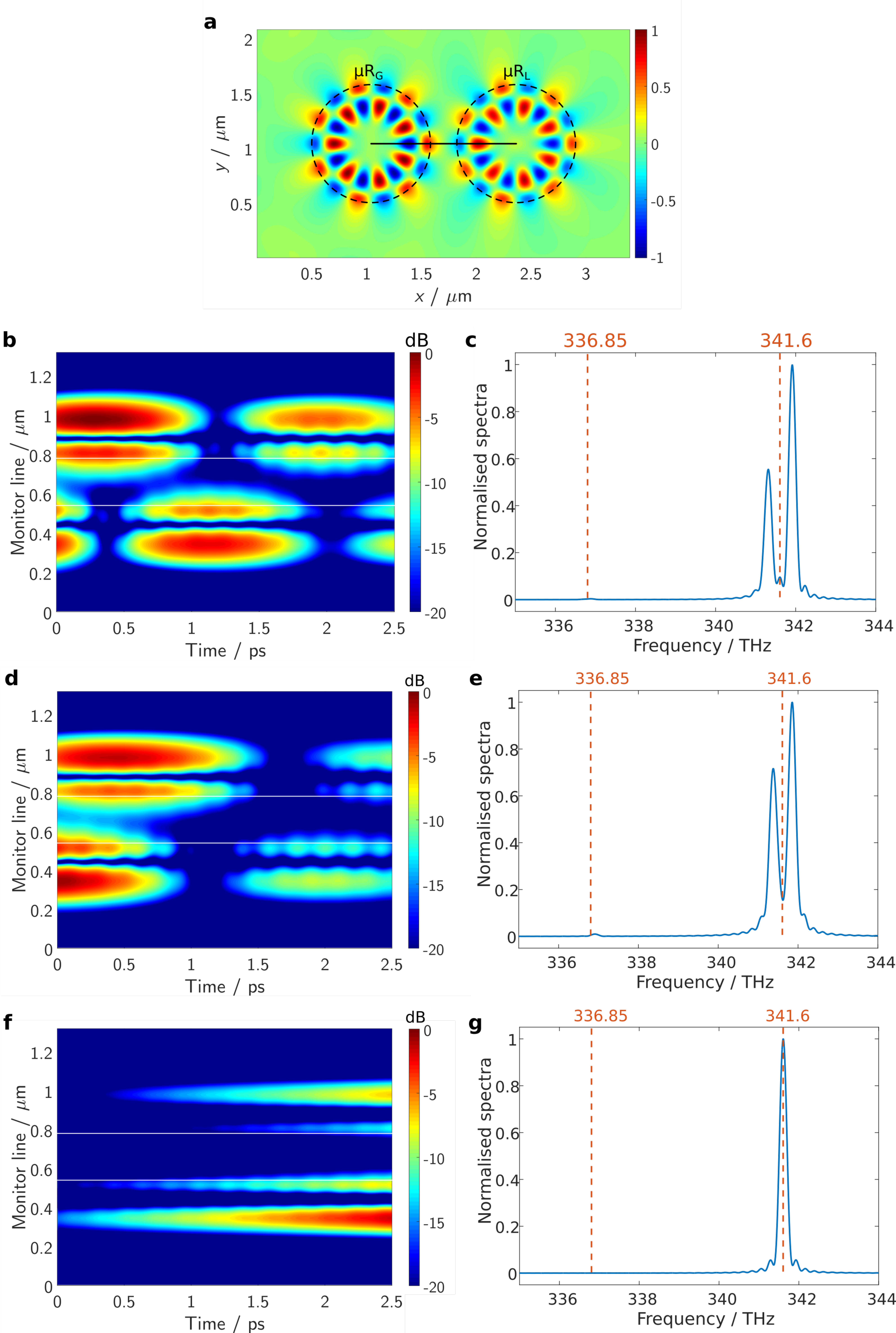}
	\end{overpic}
	\centering
	\caption[Real-time operation of \PT-microresonator coupler modelled by the 2D-TLM method.]{{Real-time operation of \PT-microresonator coupler modelled by the 2D-TLM method. ({a}) Spatial electric field distribution of the coupled microresonators operated in the (7,2) mode. The black line connecting the centre of the two resonators denotes the monitor line. The temporal evolution ({b}) and spectrum ({c}) of the field on the monitor line are shown for the passive case. (d,e,f,g) Real time operation of \PT-coupled resonators with practical dispersion parameters $\omega_\sigma\tau=212$\cite{Hagness1996} and for two different gain/loss parameter, i.e. ({d,e}) $n''(f_\text{res})=0.002$, and ({f,g}) $n''(f_\text{res})=0.0034$.}}
	\label{fig:10_tlmnogain}
\end{figure}

{Here, in Fig. \ref{fig:10_tlmnogain} we present the real-time operation of the \PT-resonator coupler only when practical material parameter is considered, $\omega_\sigma\tau=212$, for different levels of gain/loss. The spatial mode profile of a coupled WGM(7,2) is shown in Fig. \ref{fig:10_tlmnogain}(a). In this figure the black line connecting the centre of the two resonators denotes a monitor line on which the electric field is observed during the TLM simulation. Figures \ref{fig:10_tlmnogain}(b and c) show the temporal evolution and the spectra of the electric field observed along the monitor line for the case of no gain and loss, respectively.  }

{The passive case, in Fig. \ref{fig:10_tlmnogain}(b), shows a regular beating pattern in which maximum intensity is observed in one microresonator while minimum intensity is observed in the other. Slight modulation in the beating profile pattern is due to the unintentional excitation of the higher $Q$-factor WGM(10,1) mode. It is also noted that the electric field is decaying due to the radiation losses implied by the low $Q$-factor of this mode. The spectrum, in Fig. \ref{fig:10_tlmnogain}(c), shows the resonant frequencies of the super-modes which are centred around $f_\text{res}^{(7,2)}$, in agreement with Fig. \ref{fig:10_spliting}(a).}

{The temporal dynamic for the case of $n''(f_\text{res})=0.002$ (operation below threshold point) is shown in Fig. \ref{fig:10_tlmnogain}(d,e) and operation above the threshold point  $n''(f_\text{res})=0.0034$ in Fig. \ref{fig:10_tlmnogain}(f,g). In comparison to the passive case (Fig. \ref{fig:10_tlmnogain}(b)), Fig. \ref{fig:10_tlmnogain}(d) shows that the temporal dynamic for operation below the threshold point has a faster decay; this is in agreement with Fig. \ref{fig:10_spliting}(d) which shows that the imaginary part of the eigenfrequency skewed towards an overall high loss. Moreover, unlike the passive case, the coupling of the \PT-microresonator coupler for operation below the threshold  point is not regular, in a such a way that a maximum intensity at one microresonator does \textit{not} imply a minimum on another. The spectrum, which is presented in Fig. \ref{fig:10_tlmnogain}(e), shows the splitting in the resonant frequency has narrowed. The operation with gain/loss $n''(f_\text{res})=0.0034$ above the threshold point is depicted in  Fig. \ref{fig:10_tlmnogain}(f,g). The temporal dynamic of the electric field shows an exponentially increasing profile with no beating pattern observed, unlike the operation below the threshold point. The spectrum of the super-mode shows the coalesce of the two resonant frequency at $f_\text{res}^{(7,2)}$ which is also in agreement with the frequency-domain calculation performed by the BIE.    }

\section{\PT-Microresonator Photonic Molecules Array}
{In the previous section, the Boundary Integral Equation (BIE) has been developed and used to model a \PT-microresonator coupler. The impact of practical of material dispersion to the \PT-threshold point has been investigated. In this section, the BIE model is extended further to model an  array of a \PT-microresonator photonic molecules (\PT-PhM) \index{\PT-microresonator photonic molecules (\PT-PhM)} in the presence of a defect molecule.} 

{Using the BIE model, an array of \PT-PhM with a quadruplet unit cell has been studied in \cite{Phang2015d}. The report \cite{Phang2015d} shows the existence of a pair of unique \textit{non-degenerate} termination modes in a finite \PT-PhM array with spatial modulation. This pair of termination modes is unique in such a way that these modes are localised at each end of the array and their eigenfrequencies are complex-conjugates in nature, i.e. one is amplifying and the other is dissipative at the same (real) rate. For a detailed study on the \textit{defect-less} array of \PT-PhM, we refer reader to \cite{Phang2015d}; in this section we investigate the spectra of a \PT-PhM array containing a defect. In particular we report the existence of a localised defect state which is conservative in nature, i.e. which has a completely real spectrum.} 

\begin{figure}[t]
	\begin{overpic}[width=0.95\textwidth,tics=5]{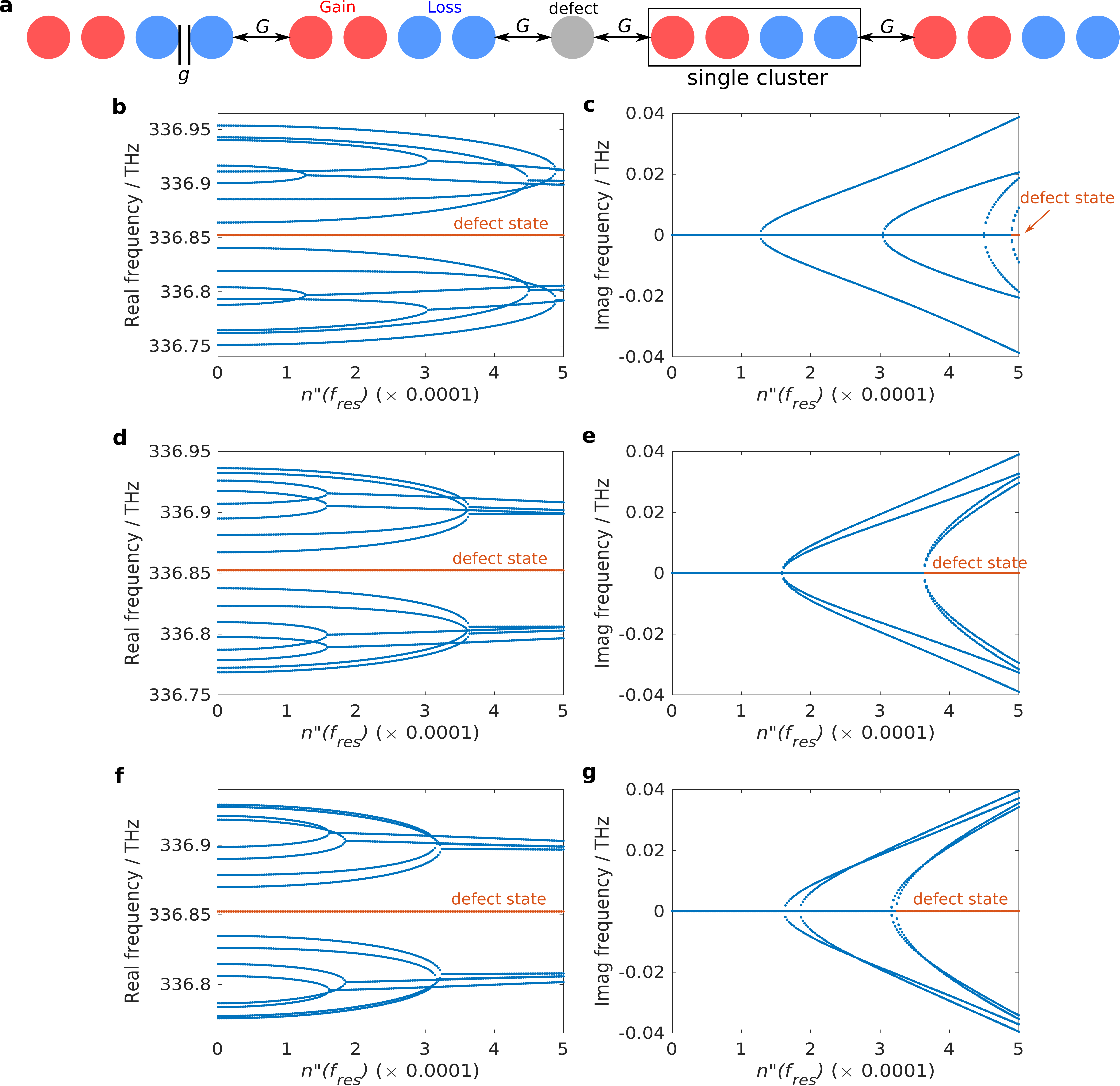}
	\end{overpic}
	\centering
	\caption[Schematic of \PT-symmetric microresonators.]{{\PT-microresonator photonic molecules (\PT-PhMs) array. (a) Schematic of \PT-PhMs array.  (b,c) Real and imaginary part of the eigenfrequencies for $G=0.9g$. (d,e) Real and imaginary part of the eigenfrequencies for $G=g$. (f,g) Real and imaginary part of the eigenfrequencies for $G=1.1g$ }}
	\label{fig:ptmolecules}
\end{figure}

{Figure \ref{fig:ptmolecules}(a) illustrates schematically the \PT-PhM structure considered in this section. The finite \PT-PhM array chain consists of four \PT-PhM clusters and a single defect located at the centre of the \PT-PhM. Each \PT-PhM cluster (inside a box in this figure) is comprised of four photonic molecules with an odd-function profile of gain/loss satisfying the \PT-symmetric condition. Each molecule within the cluster is separated by a uniform gap distance $g$ whilst distance $G$ separates the clusters and the clusters from the defect molecule. The specification of each of the microresonator molecules investigated in this section has the same geometrical and material parameters as in Section \ref{sec:ptcoup}; coupling between high $Q$-factor WGMs (10,1) is considered.}  

{The eigenfrequencies of the \PT-PhM array are presented in Fig. \ref{fig:ptmolecules}(b-g), the left-side and right-side are the real part and the imaginary part of the eigenfrequency respectively. In all cases the gap $g$ is kept constant at $0.3$ $\mu$m whilst the gap $G$ differs, i.e. $G=0.9g$, $1g$ and $1.1g$ for Fig. \ref{fig:ptmolecules}(b,c), Fig. \ref{fig:ptmolecules}(d,e) and Fig. \ref{fig:ptmolecules}(f,g), respectively. }

\begin{figure}[t]
	\begin{overpic}[width=0.9\textwidth,tics=5]{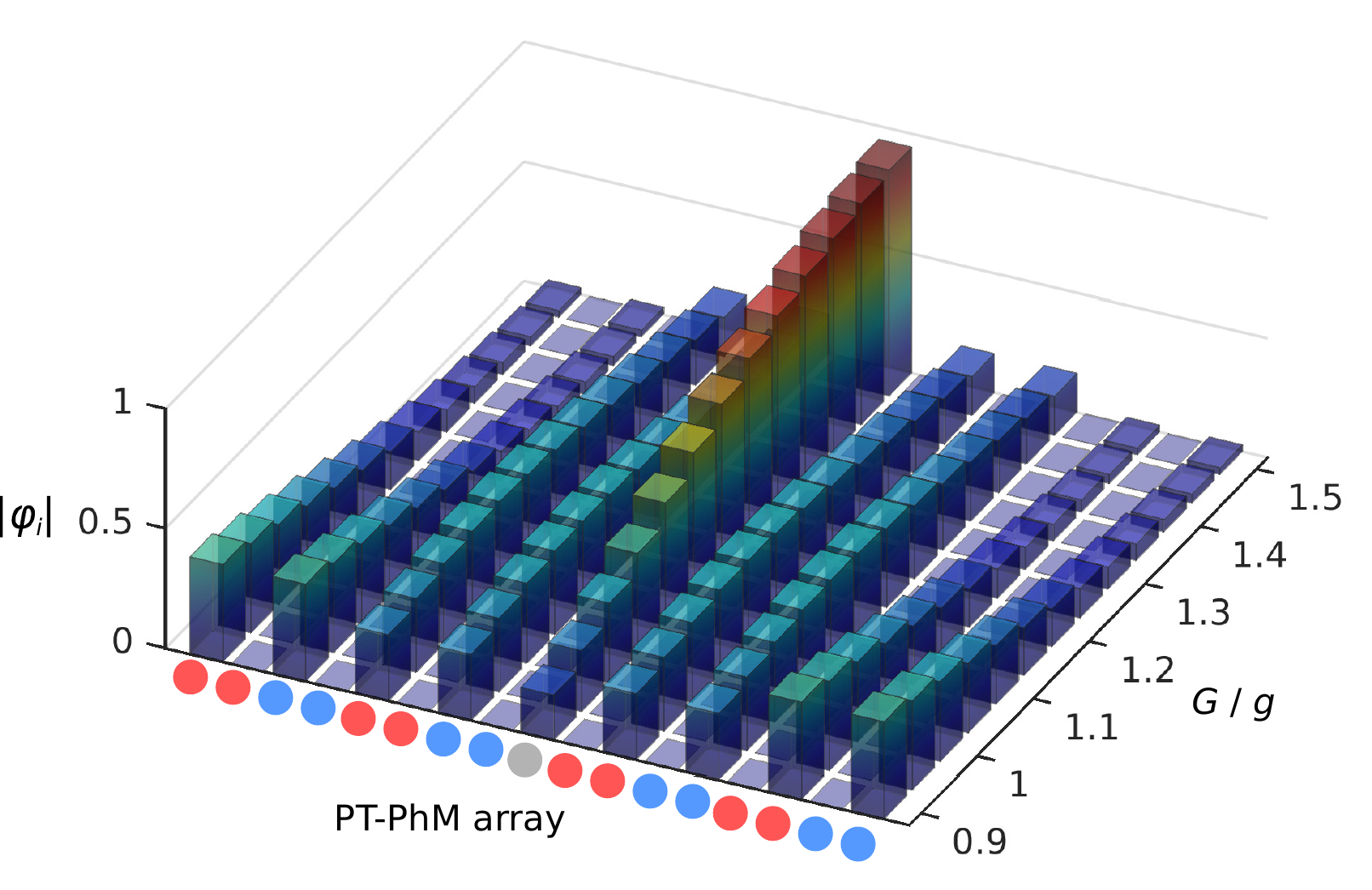}
	\end{overpic}
	\centering
	\caption[Magnitude of the eigenvector of the defect state.]{{Magnitude of the eigenvector of the defect state.}}
	\label{fig:eigvec_func_G}
\end{figure}

{Figure \ref{fig:ptmolecules}(b-g) shows a more complicated eigenspectra structure in comparison to the \PT-microresonator coupler investigated in the previous section. It is due to the fact that there are in total 17 PhMs, such that there exists 17 eigenfrequencies, which may be \textit{degenerate}. Any degenerate eigenfrequencies, in the presence of gain/loss, correspond to the \PT-threshold point at which \PT-symmetry is broken and the system operates in an unstable regime. By inspecting the imaginary part of the eigenfrequencies, i.e. Fig. \ref{fig:ptmolecules}(c,e,g), the \PT-threshold points (at which the imaginary part of eigenfrequencies splits) occur at a lower level of gain/loss for $G<g$ and at higher level of gain/loss for $G>g$. One particular eigenfrequency at $f_\text{res}^{(10,1)}$, red coloured in Fig. \ref{fig:ptmolecules}(b-g), has been denoted as a \textit{defect state}; \index{defect state}this nomenclature will be discussed below. This defect state has unique properties which differ from the other eigenfrequencies: (i) this defect state remains constants at $f_\text{res}^{(10,1)}$ regardless of the gap $G$, (ii) unlike the other modes which exist as a pair and coalesce as gain/loss increases, the defect mode has no pair and (iii) it is also completely real.}

{To further understand the defect state, Fig. \ref{fig:eigvec_func_G} shows the magnitude of the (normalised) eigenvector of the BIE matrix of the \PT-PhMs array chain operating at the defect state for different gaps $G$. The eigenvector of the BIE matrix provides the degree of excitation of the corresponding whispering gallery mode in an individual PhM. In contrast to the eigenvector operated at the other state (this result is not shown here), which is well distributed over all the PhMs, the defect state has particular localisation features, in such a way that for $G>g$ the mode exists mainly in the defect molecules and its closest neighbours. We note that this defect mode resembles topologically protected modes which exist as a consequence of time symmetry breaking\cite{Poli2015}. }

\section{Concluding remarks}
\label{sec:conc}

In this chapter a Boundary Integral Equation (BIE) method for coupled resonator was developed. The results for the BIE model were compared with the results obtained from the time-domain 2D-TLM model. It was shown that although the TLM method suffers from the red-shifting error, this error can be minimised by reducing the mesh-size or by using an enhanced TLM model incorporating unstructured meshing \cite{Sewell2004} to reduced the stair-casing error. 

The results show that the eigenfrequencies of a \PT-coupled resonators are always complex; this is a deviation of the strict definition of the \PT-symmetry which requires a balanced gain/loss. If dispersion of the gain/loss material is considered, the \PT-like behaviour is only observed at a single frequency, i.e. which matches the gain/loss atomic transitional frequency and the resonant frequency of the isolated resonator. {The real time operation of the structure is demonstrated for a practical scenario  by using the time-domain Transmission-Line Modelling (TLM) method.}  

{We also demonstrate the application of the BIE formulation to model a \PT-symmetric Photonic Molecules (PhMs) array chain. Results show that the \PT-PhMs chain has more intricate eigenspectra. We noted the presence of a defect mode which is highly localised in the defect with a completely real spectrum. }
   
It is noted that the BIE model developed in this chapter is expandable to various coupled system configurations. For example, it can be extended to study the spectral properties of a chain of resonators under \PT-symmetry as in \cite{Phang2015d} or other structures involving the coupling of resonant structures, for instance, Fano-type resonances and disordered lattice within the \PT-symmetry context.  

\bibliographystyle{spphys_rev}
\bibliography{references}



\printindex
\end{document}